\documentclass[12pt]{article}

\usepackage{amsmath}
\usepackage{amssymb}
\usepackage{euscript}

\usepackage{epsfig}

\usepackage{cite}

\usepackage{alltt}

\def\Order#1{${\cal O}(#1$)}
\def\oal{${\cal O}(\alpha)$}
\def\lint{\int\limits}
\def\bbeta{\bar{\beta}}

\newcommand{\Rcal}{{\cal R}}
\newcommand{\Bcal}{{\cal B}}
\newcommand{\koralw}{{\tt KoralW}}
\newcommand{\korwan}{{\tt KorWan}}
\newcommand{\yfsww}{{\tt YFSWW3}}
\newcommand{\photos}{{\tt PHOTOS}}
\newcommand{\tauola}{{\tt TAUOLA}}
\newcommand{\jetset}{{\tt JETSET}}
\newcommand{\tgc}{{\rm TGC}}
\newcommand{\born}{{\rm Born}}
\newcommand{\Rpar}{\Rcal_{\wp}}

\begin{document}                     

\allowdisplaybreaks

\begin{titlepage}

\begin{flushright}
{\bf  CERN-TH/2001-017\\ DESY 01-027\\ UTHEP-01-0101
}
\end{flushright}

\vspace{1mm}
\begin{center}
{\LARGE
The Monte Carlo Event Generator\\
{\tt\LARGE YFSWW3} version {\tt\LARGE 1.16}\\
for $W$-Pair Production and Decay\\
at LEP2/LC Energies$^{\dag}$
}
\end{center}

\vspace{1mm}
\begin{center}
{\bf S. Jadach$^{a,b,c}$, W. P\l{}aczek$^{d,c}$, M. Skrzypek$^{b,c}$ 
        B.F.L. Ward$^{e,c,f}$} {\em and} {\bf Z. W\c{a}s$^{b,c}$}

\vspace{1mm}
{\em
$^a$DESY Zeuthen, Platanenallee 6, D-15738 Zeuthen, Germany\\
$^b$Institute of Nuclear Physics,
  ul. Kawiory 26a, 30-055 Cracow, Poland,\\
$^c$CERN, CH-1211 Geneva 23, Switzerland,\\
$^d$Institute of Computer Science, Jagellonian University,\\
   ul. Nawojki 11, 30-072 Cracow, Poland,\\
$^e$Department of Physics and Astronomy,\\
  The University of Tennessee, Knoxville, TN 37996-1200, USA,\\
$^f$SLAC, Stanford University, Stanford, CA 94309, USA.
}
\end{center}

\vspace{1mm}
\begin{abstract}
We present the Monte Carlo event generator \yfsww\ version {\tt 1.16} 
for the process of $W$-pair production and decay in electron--positron 
collisions. It includes ${\cal O}(\alpha)$ electroweak radiative corrections in
the $WW$ production stage together with 
the ${\cal O}(\alpha^3)$ initial-state-radiation (ISR) corrections 
in the leading-logarithmic (LL) approximation,
implemented within the Yennie--Frautschi--Suura (YFS) exclusive exponentiation
framework. The photon radiation in the $W$ decays is generated by the dedicated program
\photos\ up to ${\cal O}(\alpha^2)$ LL, normalized to the $W$ branching ratios.
The program is interfaced with the $\tau$ decay library \tauola\ and
the quark fragmentation/hadronization package \jetset.
The semi-analytical code \korwan\ for the calculations of the differential and 
total cross-sections at the Born level and in the ISR approximation is 
included.
\end{abstract}

\begin{center}
{\it To be submitted to Computer Physics Communications}
\end{center}

\vspace{2mm}
\footnoterule
\noindent
{\footnotesize
\begin{itemize}
\item[${\dag}$]
  Work partly supported by 
  the Polish Government grant KBN 5P03B12420,
  the European Commission 5-th framework contract HPRN-CT-2000-00149,
  the US DoE Contracts DE-FG05-91ER40627 and DE-AC03-76ER00515,
  and the Polish--French Collaboration within IN2P3 through LAPP Annecy.
\end{itemize}
}

\vspace{1mm}
\begin{flushleft}
{\bf CERN-TH/2001-017\\ DESY 01-027
\\  January 2001
}
\end{flushleft}

\end{titlepage}

\tableofcontents
\newpage

\noindent{\bf PROGRAM SUMMARY}

\vspace{10pt}
\noindent{\sl Title of the program:} \yfsww, version {\tt 1.16}\ .

\vskip 0.2cm
\noindent{\sl Computer:}
any with the FORTRAN 77 compiler and the UNIX/Linux operating system

\vskip 0.2cm
\noindent{\sl Operating system:}
UNIX/Linux

\vskip 0.2cm
\noindent{\sl Programming language used:}
FORTRAN 77

\vskip 0.2cm
\noindent{\sl High speed storage required:}  $<$ 10~MB

\vskip 0.2cm
\noindent{\sl Size of distributed program:} 3,237~kB 

\vskip 0.2cm
\noindent{\sl Distribution format:} tar gzip file (size of 1,103~kB)

\vskip 0.2cm
\noindent{\sl No. of cards in combined program and test deck:}
about 23,000 plus about 48,000 of auxiliary packages:
\korwan, {\tt FF}, \jetset, \tauola, \photos.

\vskip 0.2cm
\noindent{\sl Keywords:}\\
Standard Model (SM), LEP2, linear colliders (LC),
quantum electrodynamics (QED), quantum chromodynamics (QCD),
boson $W$, $W$-pair production, $W$ decay, $W$ branching ratio (BR), 
triple gauge boson couplings (TGC), quartic gauge boson couplings (QGC), 
four-fermion ($4f$) background, 
radiative corrections, Yennie--Frautschi--Suura (YFS) exponentiation,
initial-state radiation (ISR), leading-log (LL) approximation, Coulomb effect, 
final-state radiation (FSR), electroweak (EW) corrections,
leading-pole approximation (LPA), Monte Carlo (MC) simulation/generation.

\vskip 0.2cm
\noindent{\sl Nature of the physical problem:}\\
The process of the $W$-pair production is important for precise 
tests of the Standard Model as well as searches for ``new physics''
at LEP2 and future linear colliders.
In order to match the experimental precision necessary for a successful
physics programme, quantum effects (the so-called radiative corrections)
have to be included into a theoretical description of this process.
It turns out that not only the so-called universal corrections 
(initial-state radiation, the Coulomb effect, ``naive'' QCD corrections, etc.)
are necessary, but also the \oal\ electroweak corrections in the $WW$
production are needed to reach the desired theoretical accuracy.  
All these effects should, preferably, be included in a Monte Carlo
event generator in order to account for realistic experimental set-ups.

\vskip 0.2cm
\noindent{\sl Method of solution:}\\
The Monte Carlo event generator for the combined $W$-pair production and decay
process including \Order{\alpha^3} LL ISR effects, the Coulomb correction
(usual or screened), the ``naive'' QCD effect, the \oal\ EW corrections
in the $WW$ production stage, implemented within the YFS exclusive exponentiation
framework is provided. 
Multiphoton radiation in the $WW$ production is generated according to 
the YFS MC method.
The photon radiation in the $W$ decays, normalized to the $W$ BRs,
is generated by the LL-type MC program \photos\ (up to two photons). 
The decays of $\tau$'s including radiative 
corrections are simulated by the dedicated
package \tauola. The quark fragmentation/hadronization is performed with the
help of the Lund program \jetset. 
The program can provide both weighted and unweighted (weight $=1$) events.
Any experimental cut and apparatus efficiency may be introduced
easily by rejecting some  of the generated events.

\vskip 0.2cm
\noindent{\sl Restrictions on the complexity of the problem:}\\
The LPA is used (in two versions) 
to describe the signal $WW$ production process. 
Multiphoton radiation according to the YFS exponentiation scheme 
is generated only for the $WW$ production stage.
\oal\ EW corrections (in LPA) are included only in the $WW$ production stage. 
The ISR effects beyond \oal\ are included in the LL approximation.
Non-factorizable corrections (interferences between the production and decay 
stages) are approximated by the so-called screened Coulomb ansatz. Spin
correlations between the $WW$ production and decays are fully included
only at the Born and the ISR levels. Radiative corrections in the $W$ decays are 
included into an overall normalization through the $W$ BRs, and
the real photon radiation is generated in the LL approximation (up to two 
photons) by the program \photos.
Anomalous triple gauge boson couplings are included in the
Born-like matrix element, i.e. with the universal SM corrections only. 
Quartic gauge boson couplings are implemented according to 
the Standard Model only (no anomalous couplings).
The $\tau$ decays and quark hadronization are performed, respectively, 
with the help of the dedicated packages \tauola\ and \jetset.
No $4f$ background processes are included. 

\vskip 0.2cm
\noindent{\sl Typical running time:}\\
200 CPU seconds of a PC Intel Pentium III @ 550MHz per 1000 unweighted events, 
for the parameter settings as they are given in the demonstration program.

\newpage

\section{Introduction}

The process of $W$-pair production in electron--positron colliders is very
important for testing the Standard Model (SM) and searching for signals of
possible ``new physics''; see e.g.~Refs.~\cite{LEP2YR:1996,LEP2YR:2000}. 
One of the main goals when investigating this process at present and future
$e^+e^-$ experiments is to measure precisely 
the basic properties of the $W$ boson, 
such as its mass $M_W$ and width $\Gamma_W$. This process also 
allows for a study of the triple and quartic gauge boson couplings at the tree level,
where small deviations from the subtle SM gauge cancellations can lead to
significant effects on physical observables -- these can be signals
of ``new physics''.  

In this work we present a Monte Carlo (MC) event generator that
simulates the production and decay of the $W$ pair at $e^\pm$ colliders.
The integrated cross sections and arbitrary differential distributions
can be calculated from a series of constant-weight fully inclusive MC events.
The program embodies the Standard Model scattering matrix element with
the \Order{\alpha} radiative corrections for the doubly-resonant component
of the scattering matrix element.
The basics of the model used in the program
and selected numerical results were already presented 
in earlier publications
\cite{yfsww2:1996,yfsww3:1998,yfsww3:2000,yfsww3:2000a}.
The important components of the presented program/calculation are
the complete \Order{\alpha} radiative correction for on-shell $W$-pair
production, which are implemented according to%
\footnote{We are grateful to the authors of these works for providing us 
          with the relevant parts of the computer code.}
Refs.~\cite{fleischer:1989,kolodziej:1991,fleischer:1993,fleischer:1994},
see also Ref.~\cite{fleischer:1995}.
The present {\tt YFSWW3} MC event generator was instrumental in achieving
the new improved theoretical precision of $0.4\%$ for the total cross section
of the $WW$ production process at the highest LEP2 energies,
as a result of direct and detailed comparison with 
the {\sc RacoonWW} MC program~\cite{racoonww-pl:2000,racoonww-np:2000}
within the 2000 LEP2 MC workshop
(this conclusion was also supported by the comparison between
{\sc RacoonWW} and the calculation of Ref.~\cite{beenakker:1999}),
see Section 4.3 in Ref.~\cite{LEP2YR:2000}.

Let us briefly discuss the basic physics properties of the $W$-pair 
production and decay process.
Since the $W$'s are unstable and short-lived particles, the $W$ pairs are not
observed directly in the experiments but through their decay products:
four-fermion ($4f$) final states (which may then also decay, radiate 
gluons/photons, hadronize, etc.). As high energy charged particles are
involved in the process, one can also observe energetic radiative photons.  
So, at the parton level, one has to consider a general process:
\begin{equation}
e^+ + e^- \longrightarrow 
4 f + n\gamma, \:\: (n=0,1,2,\ldots),
\label{4f-process}
\end{equation}
where also some background (non-$WW$) processes contribute.
In a theoretical description of this process -- according to quantum field
theory -- one also has to include virtual effects, the so-called loop 
corrections. This general process is very complicated since it involves
$\sim 80$ different channels ($4f$ final states) with complex peaking behavior
in multiparticle phase space and a large number of Feynman diagrams.
Even in the massless-fermion approximation, the number of Feynman
graphs grows up from $9$--$56$ per channel at the Born level to an enormous 
$3579$--$15948$ at the one-loop level~\cite{WW-LEP2YR:1996}. 
The full one-loop calculations have not been finished yet, even for the 
simplest case (doubly plus singly $W$-resonant diagrams)~\cite{vicini:1998}. 
But even if they
existed one would be faced with problems in their numerical evaluation
in practical applications, particularly within Monte Carlo event generators
-- they would be prohibitively huge and slow.
These are the reasons why efficient approximations in the theoretical 
description of this process are necessary. These approximations should
be such that on the one hand they would include all contributions/corrections
that are necessary for the required theoretical accuracy (dependent on the 
experimental precision) and on the other hand they would be efficient enough 
for numerical computations. Given the complicated topologies of the 
($4f+ n\gamma$) final states, such calculations should be, preferably, given 
in terms of a Monte Carlo event generator that would allow one to 
simulate the process directly~\cite{WWMC-LEP2YR:1996,4f-LEP2YR:2000}. 

Our solution to this consists of two complementary Monte Carlo event 
generators: \yfsww\ and \koralw.
The latter includes the full lowest-order $e^+e^- \rightarrow 4f$ process,
but with simplified radiative corrections -- the universal ones such as 
initial-state radiation (ISR), the Coulomb effect, etc. In \yfsww, 
on the other hand, the lowest-order process is simplified -- only the doubly 
$W$-resonant contributions are taken into account, but inclusion of the 
radiative corrections in this process goes beyond the universal ones.
In the current version of \yfsww\ only those non-universal (non-leading) 
corrections are included that are necessary to achieve the theoretical 
precision for the total $WW$ cross section of $0.5\%$ required for LEP2.
In order to achieve a gauge-invariant description of the signal $WW$ process
and to employ the existing \oal\ electroweak corrections for the on-shell $W$-pair
production, we use the so-called leading-pole approximation (LPA).
A double-pole variant of the LPA (usually referred to as DPA)
was advocated in Ref.~\cite{stuart:1997} and applied in a variety of 
calculations, see for example
Refs.~\cite{racoonww-pl:2000,racoonww-np:2000,beenakker:1999,yfsww3:2000a,placzek:2000}
and Ref.~\cite{LEP2YR:2000} for a review and further references.
The important thing is that \yfsww\ and \koralw\ programs have a well
established common part, which is the doubly $W$-resonant ($WW$) process with
the same universal radiative corrections. 
This allows us to combine the results of the two programs so as to achieve the 
desired theoretical precision for $WW$ observables. 
Detailed descriptions of \koralw\ were published 
in Refs.~\cite{koralw:1996a,koralw:1996b,koralw:1999,skrzypek:2000}.
Here, we present a documentation of the program \yfsww; 
physical aspects together with some numerical results are discussed in 
Refs.~\cite{yfsww2:1996,yfsww3:1998,yfsww3:2000,yfsww3:2000a,placzek:2000}.
Possible scenarios of combining the results of the two programs are presented
in Refs.~\cite{yfsww3:2000a,placzek:2000,koralw:2001}.

The outline of the paper is as follows. 
In Section 2, we define our notation and describe the physics contents of 
the program. In Section 3, we discuss the MC algorithm. 
In Section 4, we present the structure of the program, important routines, 
etc. Details about the practical use of the program are given in Section 5.
Finally, Section 6 summarizes the paper.
Appendices contain useful technical information on the structure of 
the program, its input/output, etc. 

\newpage
\section{Physics Contents of \yfsww}

In this section we describe we define differential distributions used
to calculate the cross section and generate events for the $W$-pair production 
and decay process.
These distributions we define quite completely, because they were
not defined in every detail in the past publications.
On the other hand we shall skip the detailed discussion of the physics
models that they represent.
This will be done in a separate publication~\cite{yfsww-physics-paper:2001}. 
First, we discuss our notation. 
Then we present the master formula and its ingredients. 
And finally, we give some details
on the MC algorithms implemented in the program. 

\subsection{Notation}
\label{Notat}

\begin{figure}[!ht]
\centering
\setlength{\unitlength}{0.1mm}
\epsfig{file=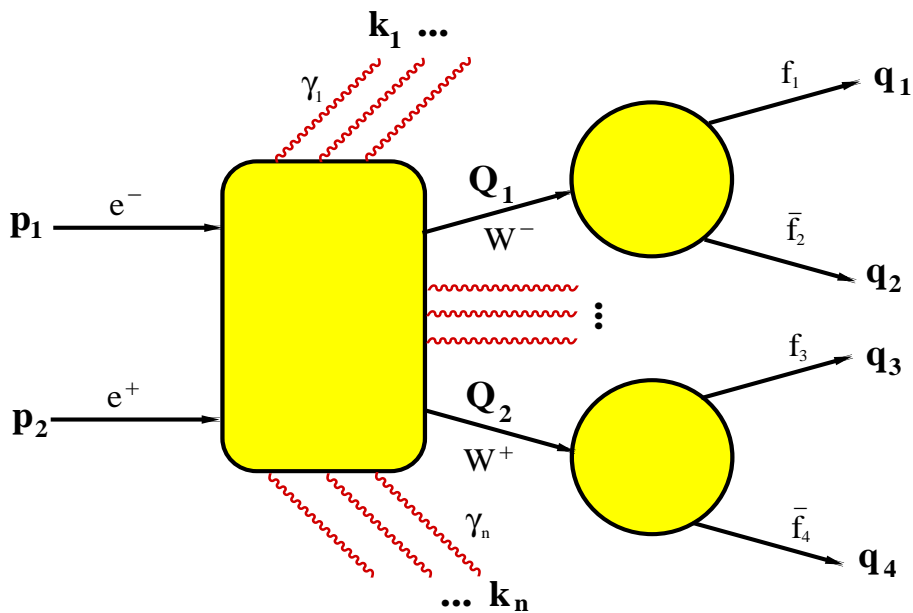,width=120mm,height=80mm}
\caption{\small\sf
  Kinematics of the $W$-pair production and decay process with multiple photons
  in the production stage.
}
\label{fig:eeWW4f}
\end{figure}

In \yfsww\ we adopt a notational convention similar to the one used
in \koralw.
The variables $p_1$ and $p_2$ are the incoming $e^-$ and $e^+$ four-momenta, 
respectively:
\begin{equation}
 p_1 = (E_e,0,0,p_e), \hspace{2cm}  p_2 = (E_e,0,0,-p_e),
\label{beams}
\end{equation}
where $E_e$ is the beam energy and $p_e=\sqrt{E_e^2 - m_e^2}$,
with the electron mass $m_e$.
The variables $Q_1$ and $Q_2$ are the four-momenta of the $W^-$ and $W^+$ bosons; 
their invariant masses are denoted by $M_1$ and $M_2$, respectively;
$q_1,q_2$ and $m_1,m_2$ are the four-momenta and masses of the $W^-$ decay 
products; $q_3,q_4$ and $m_3,m_4$ are the analogous four-momenta and masses 
of the $W^+$ decay products, with the first component of each pair 
corresponding to a fermion and the second to an antifermion. 
By $k_i$ we denote the four-momenta of radiative photons.
See also fig.~\ref{fig:eeWW4f} for a pictorial representation.
The components of a four-vector are denoted by $p=(p_0,p_x,p_y,p_z)$,
with the metric convention of $\{1,-1,-1,-1\}$.
We also use the following Lorentz invariants: $s=(p_1+p_2)^2$, 
$s_1=Q_1^2=M_1^2$ and $s_2=Q_2^2=M_2^2$.

\subsection{Master Formula}
\label{MastFor}

Here, we consider the process:
\begin{equation}
e^- + e^+ \longrightarrow 
\left\{W^- \rightarrow  f_1 + \bar{f}_2 \right\} +
\left\{W^+ \rightarrow  f_3 + \bar{f}_4 \right\}
+ n\gamma, \:\: (n=0,1,2,\ldots),
\label{WW-process}
\end{equation}
where the radiative photons are emitted only at the $WW$-production stage.
The photon radiation in $W$ decays (FSR)
in this version of \yfsww\ is done, in the \Order{\alpha^2} 
leading-logarithmic (LL)
approximation,  by the dedicated program \photos~\cite{photos:1994}
-- in the same way as in \koralw. Since this is done externally 
on fully generated events (it does not change event weights),
we do not include the FSR in our discussion of the master formula.
It should be remembered that the use of \photos\ in \yfsww\ is
a temporary solution before the advent of the YFS-exponentiated MC
for $W$ decays.

As was already mentioned in the Introduction, we use the LPA to describe this
process respecting the $SU(2)_L\times U(1)$ gauge invariance. 
In \yfsww, we implemented two options of the LPA, called LPA$_a$
and LPA$_b$. The first one, which is the default (recommended)  option,
is based on the approach of Stuart~\cite{stuart:1997}, where only the
Lorentz scalar functions in the S-matrix are expanded about poles 
corresponding to unstable $W$'s. 
In the lowest order, the LPA$_a$ matrix element coincides with the so-called
CC03 (see e.g. Ref.~\cite{WW-LEP2YR:1996} for an explanation of this notation) 
matrix element in the 't~Hooft--Feynman gauge. In the following we shall
use the notation ``CC03'' in that sense.
The second option, intended for some dedicated tests, 
follows the method suggested in Ref.~\cite{WW-LEP2YR:1996},
where the whole S-matrix residuals are expanded about poles%
\footnote{We do not apply, however, the pole approximation to the phase
          space, as was done in the semi-analytical calculations of 
          Ref.~\cite{beenakker:1999}.
         };
see Refs.~\cite{yfsww3:2000a,placzek:2000} for more details.
In the following we shall describe the LPA$_a$ scheme explicitly;
LPA$_b$ can be obtained from this by taking the $W$'s on shell
(i.e. $Q_1^2,Q_2^2\rightarrow M_W^2$) everywhere, except for the Breit--Wigner
denominators and the phase-space integration. 

In the framework of the YFS exclusive exponentiation
(EEX), the cross section for the process (\ref{WW-process}) can be written as
\begin{equation}
\begin{aligned}
\sigma_{\rm Best} =
&
\sum_{n=0}^\infty \frac{1}{n!}
\lint ds_1 ds_2 
\frac{d^3 Q_1}{Q_1^0} \frac{d^3 Q_2}{Q_2^0} 
\prod_{l=1}^4 \frac{d^3 q_l}{q_l^0} \;
\prod_{i=1}^n  \frac{d^3 k_i}{k^0_i}\;
\delta^{(4)}\bigg(p_1 + p_2 - Q_1 - Q_2 -\sum_{i=1}^n k_i \bigg)
\\&
\delta^{(4)}\left(Q_1 - q_1 - q_2 \right)
\delta^{(4)}\left(Q_2 - q_3 - q_4 \right)\;
\rho^{\rm Best}_n(p_1,p_2;q_1,\ldots,q_4,k_1,\ldots,k_n),
\end{aligned}
\label{Master}
\end{equation}
where $\rho^{\rm Best}_n$ is the following fully exclusive multiphoton 
differential distribution:
\begin{equation}
\begin{aligned}
\ & 
\rho^{\rm Best}_n(p_1,p_2;q_1,\ldots,q_4,k_1,\ldots,k_n)=
  \prod_{i=1}^n \tilde{S}(p_1,p_2,Q_1,Q_2,k_i)\;\theta(k_i^0 - k_{\epsilon})\;
\\&\times
  e^{Y'(p_1,p_2,Q_1,Q_2;k_{\epsilon})}\,
  \left[\,1 + \delta_C(Q_1,Q_2,M_W,\Gamma_W)\,\right]
 \sum_{\wp\in \{I,W\}^n} p_{\wp}\, \Bcal\left(\{p,Q,q,k\}^{\Rpar}\right),
\end{aligned}
\label{eq:rho}
\end{equation}
where $\Bcal$ is the contribution from the infrared-finite YFS residuals%
\footnote{In sums like $\sum_{\wp_{i,j}=I}$ we include all
  pairs of indices $(i,j)$, so that $i\neq j$, $\wp_i=I$ and $\wp_j=I$.}:
\begin{equation}
\begin{aligned}
  \Bcal & \left(\{p,Q,q,k\}^{\Rpar}\right) = 
  \left[\,1 + \delta_{An}^{\tgc}\left(\{p,Q,q\}^{\Rpar}\right)\,\right]\;
\\ \times & 
  \Bigg\{
  \bbeta^{(1)}_{0}\left(\{p,Q,q\}^{\Rpar}\right)
  +\sum_{i=1}^{n} 
   \frac{\bbeta^{(1)}_{1}\left(\{p,Q,q,k_i\}^{\Rpar}\right)}
        {\tilde{S}\left(\{p,Q,k_i\}^{\Rpar}\right)}
  +\Delta\bbeta^{(3)}_{0,{\rm ISR}}(\{p,Q,q\}^{\Rpar})
\\& 
  +\sum_{\wp_i=I}
   \frac{\Delta\bbeta^{(3)}_{1,{\rm ISR}}\left(\{p,Q,q\}^{\Rpar},k_i\right)}
       {\tilde{S}_I(\{p\}^{\Rpar},k_i)}
  +\frac{1}{2!}\sum_{\wp_{i,j}=I}
  \frac{\bbeta^{(3)}_{2,{\rm ISR}} \left(\{p,Q,q\}^{\Rpar},
        k_i,k_j\right)}{\tilde{S}_I(\{p\}^{\Rpar},k_i)
        \tilde{S}_I(\{p\}^{\Rpar},k_j)}
\\& 
  +\frac{1}{3!}\sum_{\wp_{i,j,l}=I}
  \frac{\bbeta^{(3)}_{3,{\rm ISR}}\left(\{p,Q,q\}^{\Rpar},
        k_i,k_j,k_l\right)}
       {\tilde{S}_I(\{p\}^{\Rpar},k_i)\tilde{S}_I(\{p\}^{\Rpar},k_j)
        \tilde{S}_I(\{p\}^{\Rpar},k_l)}
  \Bigg\},
\end{aligned}
\label{eq:betas}
\end{equation}
where we use a short-hand notation of the type 
$\{p,Q,q\}\equiv \{p_1,p_2,Q_1,Q_2,q_1,q_2,q_3,q_4\}$.

The sum $\sum_{\wp}$ in Eq.~(\ref{eq:rho}) runs over the photon partitions,
i.e. all possible photon associations to the initial-state radiation (ISR)
or the $W$-state radiation (WSR).
For $n$ photons
we identify each of the $2^n$ partitions ${\wp}$ with a vector
${\wp}\equiv (\wp_1,\ldots,\wp_n)$, where $\wp_i=I,W$
($I$ and $W$ denote the ISR and the WSR photons, respectively).
The partition weight $p_{\wp}$
is given by
\begin{equation}
p_{\wp} = 
\frac{\prod\limits_{\wp_i=I}\tilde{S}_I(k_i)
      \prod\limits_{\wp_j=W}\tilde{S}_W(k_j)}
     {\sum\limits_{\wp}\prod\limits_{\wp_i=I}\tilde{S}_I(k_i)
      \prod\limits_{\wp_j=W}\tilde{S}_W(k_j)}
= \frac{\prod\limits_{\wp_i=I}\tilde{S}_I(k_i)
       \prod\limits_{\wp_j=W}\tilde{S}_W(k_j)}
      {\prod\limits_{i=1}^n\left(\tilde{S}_I(k_i) + \tilde{S}_W(k_i)\right)},
\label{probpar}
\end{equation}
where $\tilde{S}_I(k)$ and $\tilde{S}_W(k)$ are the real photon infrared 
(IR) factors for the ISR and the WSR, respectively:
\begin{equation}
\begin{aligned}
\ & \tilde{S}_I(p_1,p_2,k) = -\frac{\alpha}{4\pi}
  \left(\frac{p_1}{kp_1} - \frac{p_2}{kp_2}\right)^2,
\\&
 \tilde{S}_W(Q_1,Q_2,k) = -\frac{\alpha}{4\pi}
   \left(\frac{Q_1}{kQ_1} - \frac{Q_2}{kQ_2}\right)^2.
\end{aligned}
\label{sisw}
\end{equation}
It is worthwhile to stress that the partition dependence for the first two terms
proportional to $\bbeta^{(1)}_{0}$ and $\bbeta^{(1)}_{1}$ cancels out,
when truncated to \Order{\alpha} (as it should), 
thanks to the special normalization choice
for the partition weights: $\sum\limits_{\wp\in \{I,W\}^n} p_{\wp}=1$.
See also the definition of $\bbeta^{(1)}_{1}$ and the discussion on
the partition-dependent ``reduction procedure'' ${\cal R_{\wp}}$ in the 
following.
The purpose of the sum over partitions is to help the introduction 
of the LL corrections {\em beyond} the \Order{\alpha}.

In Eq.~(\ref{eq:betas}) the two components proportional to
$\bbeta^{(1)}_{0}$ and $\bbeta^{(1)}_{1}$ 
explicitly represent the complete \Order{\alpha} contribution 
to the $WW$-production process, while
the remaining $\bbeta$'s and $\Delta\bar\beta$'s
collect the second- and third-order ISR corrections 
in the LL approximation;
see Section~\ref{betas} for more details.
The dependence of the $\bbeta$ functions on the photon partitions
enters through the so-called reduction (extrapolation) procedures,
which take into account the CMS energy shift induced by the ISR;
for more details see Subsection~\ref{betas}. 

\subsubsection{YFS Infrared Factors}
The real photon IR factor
\begin{equation}
\tilde{S}(p_1,p_2,Q_1,Q_2,k) = -\frac{\alpha}{4\pi^2}
\left( \frac{p_1}{kp_1} - \frac{p_2}{kp_2} 
     - \frac{Q_1}{kQ_1} + \frac{Q_2}{kQ_2} \right)^2
\label{Sfact}
\end{equation}
includes the ISR, the WSR, and the interferences between them.

The step function
$\theta(k^0-k_{\epsilon})$, where $k_{\epsilon}=\epsilon\sqrt{s}/2$,
cuts out the singular IR region, already included to all orders in the YFS 
form factor $e^{Y'}$, where
\begin{equation}
Y'(p_1,p_2,Q_1,Q_2;k_{\epsilon}) =
 2\alpha \Re B'(p_1,p_2,Q_1,Q_2) 
        + 2\alpha \tilde{B}(p_1,p_2,Q_1,Q_2;k_{\epsilon}). 
\label{YFSformf}
\end{equation}
The IR functions $\Re B'$ and $\tilde{B}$ correspond to virtual and real 
photons, respectively, and are defined explicitly in Ref.~\cite{yfsww2:1996}.

\subsubsection{Coulomb Correction}

The correction $\delta_C$ is the Coulomb effect, included both 
(two options)
in its standard form as given in Ref.~\cite{khoze:1995}, and in terms of
the so-called screened Coulomb ansatz of Ref.~\cite{scc:1999} which
is an efficient approximation
of the non-factorizable corrections of 
Refs.~\cite{beenakker:1997,beenakker:1997b,dittmaier:1998,dittmaier:1998b}
for singly-inclusive distributions.
It reads
\begin{equation}
\delta_C^{\rm \{Std,\,Scr\}}(Q_1,Q_2,M_W,\Gamma_W) = 
 \frac{\alpha\pi}{2\beta}\left[1 - \left\{1,\, (1-\beta)^2\right\}
 \frac{2}{\pi}\arctan \left(\frac{|\kappa|^2 - p^2}{2p\Re\kappa}\right)\right],
\label{Coulomb}
\end{equation}
with
\begin{equation}
\begin{aligned}
\ &\beta = \sqrt{1 -2(s_1+s_2)/s_Q +[(s_1-s_2)/s_Q]^2},\;\;\;\;
p=\frac{1}{2}\beta\sqrt{s_Q}, \;\;\;\; s_Q = (Q_1 + Q_2)^2,
\\&
\kappa =  \sqrt{\frac{M_W}{2} \left(\sqrt{E^2+\Gamma_W^2}-E\right)}  
        -i \sqrt{\frac{M_W}{2} \left(\sqrt{E^2+\Gamma_W^2}+E\right)},\;\;
E = \frac{s_Q - 4M_W^2}{4M_W},
\end{aligned}
\label{Coul:ingr}
\end{equation}
where $M_W$ and $\Gamma_W$ are the on-shell $W$ mass and $W$ width, 
respectively.
The first term in the curly brackets in Eq.~(\ref{Coulomb}) corresponds 
to the standard (Std) correction, the second one to the screened (Scr) 
Coulomb correction.
This correction is properly matched with the YFS form factor, where
the Coulomb-like singularity appears in the virtual IR function $\Re B$. 
After matching it with the full Coulomb correction $\delta_C$, the
original YFS function $\Re B$ is replaced by $\Re B'$,
as described in Ref.~\cite{yfsww2:1996}.

\subsubsection{Anomalous Triple-Gauge Couplings}

The anomalous corrections to the triple-gauge couplings (TGC) $WWV$ 
($V=Z$,$\gamma$) are included multiplicatively through 
\begin{equation}
1 + \delta_{An}^{\tgc} =
\frac{|{\cal M}_{\born}^{An\tgc}|^2}{|{\cal M}_{\born}^{\rm SM}|^2}, 
\label{delTGC}
\end{equation}
where ${\cal M}_{\born}^{An\tgc}$ is the Born-level matrix element 
with the anomalous TGCs, while ${\cal M}_{\born}^{\rm SM}$ is the 
corresponding SM matrix element. In \yfsww\ we use the same conventions 
(routines) for the TGCs as they are given in \koralw~\cite{koralw:1999}.
The anomalous couplings are included in the event generation 
only if an appropriate switch in the input parameters is on and the
parameters corresponding to these couplings (in a given parameterization) 
are set up to non-SM values; see Appendix~A for details.

\subsubsection{$\bbeta$'s for $W$-Pair Production}
\label{betas}

In the IR-finite $\bbeta$ functions of Eq.~(\ref{eq:rho}), we see
the superscripts $\Rpar$ in the four-momenta arguments like
$\{p,Q,q\}^{\Rpar}$ or $\{p,Q,q,k\}^{\Rpar}$.
This means that these arguments are
subject to the ``reduction (extrapolation) procedures''.
These reduction procedures are partition-dependent -- this is why they
are denoted with the subscripts $\wp$.
The partition dependence enters through accounting for the ISR energy shift 
$s\rightarrow s'$, where $s'=(p_1+p_2-\sum_{\wp_i=I} k_i)^2$
in the evaluation of the $\bbeta$ functions.
For more discussion on the meaning of the reduction/extrapolation procedures,
see Subsection ~\ref{sec:reduction}.
In \yfsww, we use  the reduction procedures of the {\tt YFS2} and {\tt YFS3}
programs of {\tt KORALZ}~\cite{yfs2:1990,yfs3:1992,koralz:1994}, 
with some modifications: in constructing the reduced $W$ four-momenta, 
we take into account that, in general, $M_1\neq M_2$.
We perform appropriate transformations of the $W^\pm$ decay products, that is
we boost them to $W$'s rest frames before the reduction, and then make 
the inverse boost transformation to the reduced frame.
In general, the effect (recoil) of the CMS energy shift $s\rightarrow s'$ is taken
due to the ISR photons only.
This procedure is well justified for the ISR corrections beyond \Order{\alpha}
where we take $\{p,Q,q\}^{\Rpar}$ and for each photon we may
say whether it belongs to the ISR or to the WSR.
For the \Order{\alpha} specific corrections
we cannot make a distinction, in the corresponding $\bbeta$'s,
between the ISR and the WSR;
the $\{p,Q,q,k_i\}^{\Rpar}$ in their arguments means that
(i) the reduced $\{p,Q,k_i\}^{\Rpar}$ are within the 3-body phase space
and
(ii) the photon $k_i$ is excluded from the energy shift $s\rightarrow s'$
calculation.
For more discussion on this non-trivial point 
see Subsections \ref{sec:reduction} and \ref{sec:multi-reduction}.

Let us now describe in detail the \Order{\alpha}-complete%
\footnote{In the early version of the present program called {\tt YFSWW2},
          corrections beyond the ISR were included only through the function $Y'$ 
          and $\tilde{S}$-factors~\cite{yfsww2:1996}.}
functions $\bbeta^{(1)}_{0}$ and $\bbeta^{(1)}_{1}$.
The first one, which includes the \oal\ electroweak (EW) virtual 
corrections, reads
as follows:
\begin{equation}
  \bbeta^{(1)}_{0}(\{p,Q\}) = 
  \bbeta^{(0)}_0(\{p,Q\}) 
  \left[1 +
    \delta^{(1)}_{v+s}(\{p,Q\};k_{\epsilon})
    - Y(\{p,Q\};k_{\epsilon}) 
  \right],
  \label{beta0}
\end{equation}
where $\bbeta^{(0)}_0$ 
is the lowest-order matrix element squared
for the $W$-pair production and decay; it is identical
to the corresponding $\bbeta^{(0)}_0$ function in
\koralw~\cite{koralw:1996a,koralw:1999}.
In fact, the $\bbeta^{(1)}_{0}$ function comes in two variants.
The first one, referred to as {\em scheme (A)}~\cite{yfsww3:2000a},
reads as follows:
\begin{equation}
\begin{aligned}
  \bbeta^{(1)}_{0,A}(\{p,Q\}) &=
  \bbeta^{(0)}(\{p,Q\})\,\left[1+ \delta^{(1)}_{\rm ISR}(p_1,p_2)\right]\\
& + \bbeta^{(0)}(\{p,Q\})
  \bigg[ \delta^{(1)}_{v+s}(\{p,Q\};k_{\epsilon})
         -Y(\{p,Q\};k_{\epsilon}) 
         -\delta^{(1)}_{\rm ISR}(p_1^{\rm eff},p_2^{\rm eff}) 
  \bigg]_{G_{\mu}}
  \label{beta0A}
\end{aligned}
\end{equation}
where $\delta^{(1)}_{\rm ISR}=(\alpha/\pi)[\ln(s/m_e^2) -1]$ 
is the LL pure QED \oal\ ISR correction,
and the subscript $G_{\mu}$ means that the whole correction is calculated
in the so-called $G_{\mu}$ scheme~\cite{fleischer:1989} as explained in
Ref.~\cite{yfsww3:2000a}.
Another option, referred to as {\em scheme (B)},
similar to the one used in the program {\sc RacoonWW}~\cite{racoonww-np:2000},
is defined as follows:
\begin{equation}
\begin{aligned}
  \bbeta^{(1)}_{0,B} & (\{p,Q\}) =
  \bbeta^{(0)}(\{p,Q\})\,\left[1+ \delta^{(1)}_{\rm ISR}(p_1,p_2)\right]\\
&+\bbeta^{(0)}(\{p,Q\})
  \bigg[ \delta^{(1)}_{v+s}(\{p,Q\};k_{\epsilon})
         -Y(\{p,Q\};k_{\epsilon}) 
         -\delta^{(1)}_{\rm ISR}(p_1^{\rm eff},p_2^{\rm eff}) 
  \bigg]_{G_{\mu}} \left(\frac{\alpha}{\alpha_{G_{\mu}}}\right),
  \label{beta0B}
\end{aligned}
\end{equation}
where $\alpha$ is the fine structure constant and 
$\alpha_{G_{\mu}}=(\sqrt{2}G_{\mu} M_W^2 \sin^2\theta_W)/\pi$. 
Here, although the \oal\ corrections are calculated in the $G_{\mu}$ scheme,
their relative coefficient is $\alpha$ instead of $\alpha_{G_{\mu}}$.
Our recommended option is {\em scheme (A)}, which is inspired by
the renormalization group equations (RGEs) \cite{bennie:1987,yfsww3:2000a}.

The \oal\ virtual
and soft photon corrections for the $WW$ production stage are included in 
$\delta^{(1)}_{v+s}$. In the current version of the program, we use 
the calculations of Ref.~\cite{fleischer:1989} for the on-shell $WW$-production. 
Since in our case the $W$'s are off-shell, we have to perform some 
transformations/extrapolations 
in order to employ these 
calculations. We do it in the following way. After performing the reduction
procedure for the $\bbeta_0$ function, as described above, we find 
$\cos\theta_{\rm eff}$ and the velocity $\beta_{\rm eff}$ of the $W$'s in the reduced 
frame, where
\begin{equation}
  \begin{aligned}
   \ &\beta_{\rm eff} = 
    \sqrt{1 - 2(s_1+s_2)/s' + [(s_1-s_2)/s']^2},\\
   &\cos\theta_{\rm eff}=
     (\vec{p}_1^{\Rpar}\cdot\vec{Q}_1^{\Rpar})/(|\vec{p}_1^{\Rpar}| 
     |\vec{Q}_1^{\Rpar}|).
  \end{aligned}
  \label{betaeff}
\end{equation}
Then, for this value of $\beta_{\rm eff}$ we calculate the effective CMS energy 
squared for on-shell $W$'s:
\begin{equation}
s_{\rm eff} = 4 M_W^2/(1 - \beta_{\rm eff}^2).
\label{seff}
\end{equation}
Having ($s_{\rm eff}$, $\beta_{\rm eff}$, $\cos\theta_{\rm eff}$), we construct
the four-momenta $(p_1^{\rm eff},p_2^{\rm eff},Q_1^{\rm eff},Q_2^{\rm eff})$ for the on-shell
process 
$e^ + e^- \rightarrow W^ + W^-$ 
and use them to calculate the respective EW
corrections. The same four-momenta are used to calculate the IR form factor 
$Y$ according to Eq.~(\ref{YFSformf}); now
the Coulomb-like singularity in the virtual IR function
$\Re B$ is kept, in order to cancel the corresponding singularity 
in $\delta^{(1)}_{v+s}$, and to calculate the \oal\ ISR correction:
\begin{equation}
\delta^{(1)}_{\rm ISR}(p_1^{\rm eff},p_2^{\rm eff})=
\frac{\alpha}{\pi}\left(\ln\frac{s_{\rm eff}}{m_e^2} - 1 \right).
\label{delISR}
\end{equation}
Such a procedure allows us to use the EW corrections also in the
$WW$-threshold region, however, in this region the LPA becomes less precise.
Since there exists no other calculation with the \oal\ EW corrections for the $W$-pair
production and decay process in the $WW$-threshold region, we estimate,
for the moment, the precision of \yfsww\ in that region 
at $2\%$ conservatively, just as for the ISR approximation. This can be improved in 
the future, when more calculations/tests are done. 

Two comments are in order about the approximations used in the evaluation of
Eqs.~(\ref{beta0}--\ref{beta0B}):
\begin{enumerate}
\item
  In the LPA$_{a,b}$, the one-loop corrections should be evaluated ``on-pole'',
  i.e. for the complex $W$ mass: $M^2 = M_W^2 - iM_W\Gamma_W$.
  This would, however, require the analytic continuation of the
  on-shell one-loop corrections to the second Riemann sheet. 
  We avoid this by using the approximation 
  $M\simeq M_W$ in the LPA residuals. Since this is done for the \oal\ corrections,
  the error introduced by this approximation is of the order of
  $(\alpha/\pi)(\Gamma_W/M_W) < 10^{-4}$, which is negligible for
  the precision we aim at.
\item
  A coherent sum over $W$ spins between the production and decay stages
  is included in Eq.~(\ref{beta0}) in the $\bbeta_0^{(0)}$ function, but not
  in the non-leading (NL) \oal\ EW correction. 
  This should be sufficient for the LEP2 
  precision as the total EW correction in the LEP2 energy range is typically 
  $\sim 1$--$2\%$. However, the $W$-spin effects can be included also
  in the NL \oal\ EW correction, if necessary.
\end{enumerate}

The \oal\ real photon ISR and photon emission in the $WW$-production stage
are included through the following IR-finite function:
\begin{equation}
  \begin{aligned}
    \bbeta^{(1)}_{1}\left(\{p,Q,q,k\}^{\Rpar}\right) 
    & =  D_{\gamma}^{(1)}\left(\{p,Q,q,k\}^{\Rpar}\right) 
    - \tilde{S}\left(\{p,Q,q,k\}^{\Rpar}\right)\, 
    \bbeta_0^{(0)}\left(\{p,Q,q\}^{\Rpar}\right),
  \end{aligned}
  \label{beta1}
\end{equation}
where $D_{\gamma}^{(1)}$ is the differential cross section for
the single hard photon radiation in the $WW$-production stage
and the non-radiative $W$-decays.
For this we use the calculations of Ref.~\cite{kolodziej:1991},
where the $W$'s are taken on shell. In order to accommodate them in our 
off-shell process, we use general formulae for $W$-polarization
vectors in terms of their off-shell four-momenta \cite{hagiwara:1986} 
in the spin amplitude calculations, and perform a coherent summation
over the $W$'s polarizations of the production and decay amplitudes. 
In fact, the hard-photon amplitudes of Ref.~\cite{kolodziej:1991} are given
in the massless-fermion approximation, so they become inaccurate 
when a photon is emitted close to a fermion direction.  
To correct this, we supplement these amplitudes with the mass terms according
to the CALKUL prescription of Ref.~\cite{berends:1982}.

In Eqs.~(\ref{Master}) and (\ref{eq:rho}) we use
the IR-finite $\bbeta$ functions for the ISR
$\bbeta_{0,{\rm ISR}}^{(3)}$, $\bbeta_{1,{\rm ISR}}^{(3)}$, $\bbeta_{2,{\rm ISR}}^{(3)}$ 
and $\bbeta_{3,{\rm ISR}}^{(3)}$, calculated up to \Order{\alpha^3}~LL.
They are exactly the  same  as in \koralw, see Ref.~\cite{koralw:1999}.
We define in addition:
\begin{equation}
  \Delta\bbeta_{0,{\rm ISR}}^{(3)}=\bbeta_{0,{\rm ISR}}^{(3)}-\bbeta_{0,{\rm ISR}}^{(1)},\quad
  \Delta\bbeta_{1,{\rm ISR}}^{(3)}=\bbeta_{1,{\rm ISR}}^{(3)}-\bbeta_{1,{\rm ISR}}^{(1)}.
\end{equation}
In Eqs.~(\ref{Master}) and (\ref{eq:rho}), they are evaluated only for the ISR 
photons.
In principle, these ISR residuals could be calculated for all photons
-- it is still consistent with the LL ISR approximation.

\subsubsection{Pretabulated EW Corrections}
\label{pretab}

We have implemented also the pretabulated (or ``fast'') version of the 
EW corrections. 
In this mode, a pretabulation of the NL \oal\ EW corrections 
\begin{equation}
\delta_{\rm EW}^{\rm NL}(\{p,Q\}) = \delta^{(1)}_{v+s}(\{p,Q\};k_{\epsilon})
   -Y(\{p,Q\};k_{\epsilon}) -\delta^{(1)}_{\rm ISR}(p_1,p_2) 
\label{eq:pretab}
\end{equation}
is done in the initialization stage of the program. Then, during the event
generation, a linear interpolation of the pretabulated values is
performed for each event. 
The pretabulation is done as a function of $\cos\theta$ of the $W^-$ 
for fixed values of the CMS energy $\sqrt{s}$ (set-up to the nominal energy) 
and of the $W$ invariant masses (set-up to $M_W$). 
This means, that in this approach, the ISR energy shift $s\to s'$ as well
as the off-shellness of the $W$'s is not taken into account (the latter
prevents this method from being used below the $WW$-threshold). 
In that respect, the pretabulated version of the EW corrections should 
be regarded as some approximation of the \yfsww\ ``best'' prediction. 
The main advantage of this version is that it is much faster in computing
time than the exact (``best'') one, because here the time-consuming
EW library is invoked only at the beginning of the program run 
for some number of equally spaced points instead of by calling it for each 
generated event.  
We have checked that at $\sqrt{s}=200$~GeV, 
the results for the total cross section and main distributions 
of the pretabulated version differ from the corresponding results of 
the ``best'' one by only $0.1\%$, which is well below the required
theoretical precision for the LEP2 data analysis.

\subsubsection{Fixed-Order Calculation}

By truncating the master formula~(\ref{Master}) at some power of
the coupling constant $\alpha$, one can obtain a fixed-order calculation,
e.g. \Order{\alpha^0}, \Order{\alpha^1}.

At \Order{\alpha^0} (the Born approximation), one gets the differential 
distribution
\begin{equation}
\rho_0^{(0)}(\{p,Q,q\}) = \bbeta_0^{(0)}(\{p,Q,q\}).
\label{fixed0}
\end{equation} 

At \Order{\alpha^1}, there are two contributions: (1) from the one-loop
virtual corrections and the real soft photon radiation 
\begin{equation}
\rho_0^{(1)}(\{p,Q,q\};k_{\epsilon}) = \bbeta_0^{(0)}(\{p,Q,q\})
\left[1 + \delta^{(1)}_{v+s}(\{p,Q\};k_{\epsilon})\right],
\label{fixed1vs}
\end{equation} 
and (2) from the real hard photon radiation
\begin{equation}
\rho_1^{(1)}(\{p,Q,q\},k) = D_{\gamma}^{(1)}(\{p,Q,q\},k). 
\label{fixed1h}
\end{equation} 

We have implemented the above fixed-order calculations in \yfsww.
They are available through the parallel weights, as explained in 
Table~\ref{tab:yfsww-weights} of Appendix~A.
For $\delta^{(1)}_{v+s}$ we use the calculation of Ref.~\cite{fleischer:1989},
and for $D_{\gamma}^{(1)}$ the calculation of Ref.~\cite{kolodziej:1991};
see also Subsection~\ref{betas} for more details.

\subsubsection{Radiative Corrections for $W$ Decays}

Radiative corrections in $W$ decays are not included in the above formulae
explicitly. However, we include them in the overall normalization of
the cross section by normalizing the $W$ decay amplitudes to the $W$ branching
ratios (BR), which can be provided through the input parameters of the
program. This means that in the $\bbeta$ functions of Eq.~(\ref{eq:betas})
for the $(i,j)$-th $W^-W^+$ decay channel, we use the normalization constant 
\begin{equation}
{\cal N}_{ij} = \frac{B_i B_j}{B_{ref}^2},
\label{eq:BR}
\end{equation} 
with 
\begin{equation}
B_{ref} =\frac{\alpha_W M_W}{12\Gamma_W\sin^2\theta_W},
\label{eq:BRref}
\end{equation} 
where $\theta_W$ is the weak mixing angle, and $\alpha_W=\alpha_{G_{\mu}}$
in our recommended $G_{\mu}$ scheme, while
we set $\alpha_W=\alpha$ in the so-called $\alpha$ scheme.
$B_i$ is the BR for the $i$-th $W$-decay channel, while
$B_{ref}$ is the reference BR used for the normalization of the actual 
matrix element; see Ref.~\cite{koralw:1996a} for more details. 
For the fully inclusive process (all channels), 
the normalization constant is
\begin{equation}
{\cal N}_{all} = \frac{1}{B_{ref}^2}.
\label{eq:BRall}
\end{equation}

It turns out that the BRs calculated at the Born level in the
$G_{\mu}$ scheme -- the so-called improved Born approximation (IBA) --
agree within $0.1\%$ with the full one-loop results~\cite{WW-LEP2YR:1996}. 
Such an IBA option is also included in the program -- for this one needs
only to supply (through the input parameters) the values of the CKM matrix 
elements. In that case
\begin{equation}
B_i =
\left\{
  \begin{array}{l}
    B_{ref}:  \hspace{48mm} {\rm for \:\; leptons,}\\
    3\,|V_i^{\rm CKM}|^2 \left(1+\frac{\alpha_s(M_W^2)}{\pi}\right)
    B_{ref}: \hspace{4mm}{\rm for \:\; quarks,}
  \end{array}
\right.
\label{eq:BRgmu}
\end{equation}
where $V_i^{\rm CKM}$ is the CKM matrix element for the $i$-th $W$-decay
channel, and $\alpha_s(M_W^2)$ is the QCD coupling constant at the $M_W$
scale.

By construction, the real photon radiation in $W$ decays, which is generated by
\photos, does not affect the overall normalization of the cross section --
this is of course motivated by the KLN theorem~\cite{kinoshita:1962,lee-nau:1964}.

\subsubsection{ISR Reference Differential Distribution}

Equations~(\ref{Master}) and (\ref{eq:rho}) define the so-called ``best'' physics 
model of \yfsww; if the corresponding weight is used as a rejection
weight in order to generate unweighted (constant-weight) events, 
then we say that \yfsww\ is run in the ``best'' mode.
We want to combine the fully exclusive (multiphoton) differential
distributions from \yfsww\ and \koralw, so that the resulting distributions
contain both the LPA \oal\ EW corrections (missing in \koralw) 
and $4f$-background diagram corrections (missing in \yfsww), 
using event-per-event correction weights.
In order to do this, we must define a certain simplified, auxiliary,
fully exclusive, differential distribution $d\sigma_R$, which is
{\em common to both programs},
and implement the corresponding MC weight $w_R$ in both programs.
In \yfsww\ the weight $w_R$ can be used also optionally, as a main rejection 
weight -- this is the so-called ``ISR mode''.
Otherwise, $w_R$ is one of many alternative weights calculated
simultaneously with the ``best model weight''; it is available to the user,
and/or it is used to calculate the correction weight that
serves the purpose of
introducing the missing \oal\ EW corrections in events generated by \koralw.

In the reference distribution $d\sigma_R$, only the {\em numerically leading} 
corrections, i.e. the ISR, the Coulomb correction, etc., are taken into 
account. This reference distribution
includes the CC03 matrix element defined identically in \yfsww\ and \koralw.
We also say, sometimes, that this is the distribution of the ISR-type model.
The corresponding fully exclusive, reference, differential distribution,
replacing $\rho_n$ of Eq.~(\ref{eq:rho}),
we define as follows:
\begin{equation}
\begin{aligned}
\ &\rho^{R}_{n}(p_1,p_2;q_1,\ldots,q_4,k_1,\ldots,k_{n})=
  \prod_{i=1}^{n} \tilde{S}_I(p_1,p_2,k_i)\;\theta(k_i^0 - k_{\epsilon})\;
\\&\times
  e^{Y_I(p_1,p_2;\epsilon)}\;
  \left[\,1 + \delta_C(Q_1,Q_2,M_W,\Gamma_W)\,\right]
  \left[\,1 + \delta_{An}^{\tgc}\left(\{p,Q,q\}^{\Rcal}\right)\,\right]\;
\\& 
  \Bigg\{
  \bbeta^{(3)}_{0,{\rm ISR}}\left(\{p,Q,q\}^{\Rcal}\right)
  +\sum_{i=1}^{n} 
   \frac{\bbeta^{(3)}_{1,{\rm ISR}}\left(\{p,Q,q\}^{\Rcal},k_i\right)}
        {\tilde{S}_I\left(\{p\}^{\Rcal},k_i\right)}
  +\sum_{i>j}^{n}
  \frac{\bbeta^{(3)}_{2,{\rm ISR}} \left(\{p,Q,q\}^{\Rcal},
        k_i,k_j\right)}{\tilde{S}_I(\{p\}^{\Rcal},k_i)
        \tilde{S}_I(\{p\}^{\Rcal},k_j)}
\\& 
  +\sum_{i>j>l}^{n}
  \frac{\bbeta^{(3)}_{3,{\rm ISR}}\left(\{p,Q,q\}^{\Rcal},
        k_i,k_j,k_l\right)}
       {\tilde{S}_I(\{p\}^{\Rcal},k_i)\tilde{S}_I(\{p\}^{\Rcal},k_j)
        \tilde{S}_I(\{p\}^{\Rcal},k_l)}
  \Bigg\},
\end{aligned}
\label{eq:rhoISR}
\end{equation}
where we have made the following 
replacements with respect to Eq.~(\ref{eq:rho}):
\begin{equation}
  \begin{aligned}
   \ & Y'\;\longrightarrow \;
      Y_I(p_1,p_2;\epsilon) = 2\frac{\alpha}{\pi}
            \left[\left(\ln\frac{s}{m_e^2} - 1 \right)\ln\epsilon 
            + \frac{1}{2}\ln\frac{s}{m_e^2} - 1  
            + \frac{\pi^2}{2}\right],           
   \\& \tilde{S}\;\longrightarrow \;
       \tilde{S}_I(p_1,p_2,k), 
  \end{aligned}
  \label{YS-ISR}
\end{equation}
and we include the pure ISR LL $\bbeta$'s only.
The WSR is just absent from the above distribution,
which means that there is only one photon partition (where all photons
are associated to the ISR) with the weight $p_{\wp}=1$.
In the actual MC program, $\rho^{R}_{n}$ should be
implemented with a MC weight in the operational mode when the WSR is 
switched off -- through the parallel weights.
Both of these possibilities are included in \yfsww.
However, it can also be implemented as a MC weight when the WSR 
is switched on.
In such a case, the sums over the $\bbeta$'s should extend to all photons.

The above ISR reference cross section is also used
in this and other works to define the so-called non-leading (NL) corrections
\begin{equation}
\sigma_{\rm NL} \equiv \sigma-\sigma_{R} = \int \left(d\sigma - d\sigma_{R}\right).
\end{equation}
The subscript NL expresses the fact that these corrections 
are numerically smaller than the leading ISR corrections. 
It does not mean, however, that these corrections are free of logarithmic 
contributions, such as $\ln(s/M_W^2)$, 
but these logarithms are numerically much smaller than the ISR 
basic logarithm $\ln(s/m_e^2)$.

As already indicated, the reference ISR differential distribution
of Eq.~(\ref{eq:rhoISR})
(for the same CC03 matrix element) is also implemented in \koralw, 
and it has been checked numerically to a very high 
accuracy~\cite{yfsww3:2000a,koralw:2001}
that its \koralw\ implementation 
agrees with that of the present \yfsww\ program.
Having done that, it is possible
to combine the results of the \koralw\ and \yfsww\ programs by reweighting 
the MC events, in order to include both the effects of 
the $4f$ background and the LPA \oal\ NL corrections in the MC predictions
for the $W$-pair production and decay process.
This is done in practice by reweighting the MC events produced by \koralw\
with the correction weight provided by \yfsww.
The correction $\delta^R_{\rm NL}$ to be used by \koralw\ is defined as
\begin{equation}
  1+\delta^R_{\rm NL} = 
  \frac{\rho^{\rm Best}_{n}(p_1,p_2;q_1,\ldots,q_4,k_1,\ldots,k_{n})}
       {\rho^{R}_{n}(       p_1,p_2;q_1,\ldots,q_4,k_1,\ldots,k_{n})},
\label{eq:deltaNL}
\end{equation}
where the numerator and denominators are defined in Eqs.~(\ref{eq:rho}) 
and (\ref{eq:rhoISR}).
A few remarks are in due order.
The above ratio is, in fact, calculated as the ratio of the corresponding 
MC weights.
The \yfsww\ program has to be run in the mode with the WSR switched on, 
because $\rho^{\rm Best}_{n}$ includes the summation over 
the ISR--WSR photon partitions, see Eq.~(\ref{eq:rho}),
through the multichannel reduction procedure $\Rpar$.
Under these conditions the above correction includes the complete \oal\
correction, which can be used to correct the fully exclusive 
distributions generated by \koralw, on an event-per-event basis.
As we shall see in Subsection~\ref{rewt},
the \yfsww\ program provides at present only a
numerically efficient approximation of the above $\delta^R_{\rm NL}$.

\subsection{YFS Reduction/Extrapolation Procedures}
\label{sec:reduction}

It is characteristic and inherent feature of the YFS exponentiation that
the IR-finite $\bbeta_i$ distributions require extrapolation to a larger
phase space with the additional ``spectator'' photons.
For instance, the function $\bbeta_0$ is in our case originally defined at 
the $(\{q_1,q_2,q_3,q_4\})=(\{ q\})$ point of the 4-particle phase space
and it has to be also defined at every $(\{q\},k_1,\ldots,k_n)$ point
of the ($4+n$)-particle phase space.
We call the procedure of extending the domain of the $\bbeta_0$ function
$\bbeta_0(\{q\}) \to \bbeta^E_0(\{q\},k_1,\ldots,k_n)$ 
the ``extrapolation procedure''.
The only true limitation in the choice of the extrapolation procedure
is that $\bbeta^E_0(\{ q\},0,\ldots,0)=\bbeta_0(\{q\})$ must hold.
Typically, $\bbeta_0(\{q\})$ is defined in terms of dot-products
of the four-momenta $p_i\cdot q_j$ (or inner spinor products).
Extrapolation may be done in a natural way by using exactly the same algebraic
expressions in the larger phase space as were originally obtained 
from the Feynman diagrams in the smaller phase space.
In the smaller phase space, the dot products obey certain relations due to 
the four-momentum conservation,
which no longer hold true in the presence of the additional photons
-- this is the main mechanism in the realization of such an extrapolation.
Another useful method of implementing the extrapolation 
(especially in the case when $\bbeta_0(\{q\})$ is implemented in the form of 
a black-box procedure, which requires as the  input the four-momenta 
in the 4-particle phase space) is to project the $(\{q\},k_1,\ldots,k_n)$ 
point into the ``reduced'' point $\{q\}^{\Rcal}$
using some kinematical manipulations (boosts, rotations, rescalings) 
on the four-momenta, and subsequently plugging it into 
$\bbeta^E_0(\{q\},k_1,\ldots,k_n) \equiv \bbeta_0(\{q\}^{\Rcal})$.
The above ``reduction procedure'' is, of course, less general than the 
extrapolation procedure.
Last, but not least, let us note that the extrapolation procedure 
$(\{q\}) \to (\{q\},k_1,\ldots,k_n)$ is the reverse procedure
to that of defining the residua at the IR-divergent points/poles
$(\{q\},k_1,\ldots,k_n) \to (\{q\})=(\{q\},0,\ldots,0)$ 
in the derivation of the resummation of the IR divergences
to the infinite order (i.e. the derivation of the YFS exponentiation).
The extrapolation procedure
$\bbeta_1(\{q\},k_j) \to \bbeta^E_1(\{q\},k_1,\ldots,k_j,\ldots,k_n)$ is quite 
analogous.
It has to obey the relation 
$\bbeta_1(\{q\},k_j) \to \bbeta^E_1(\{q\},0,\ldots,0,k_j,0,\ldots,0)$.
Here all photons except the $j$-th one are the spectators.

It should be stressed that the above freedom, due to
a reduction/extrapolation procedure, is a well-known feature of the YFS exponentiation,
already underlined in Ref.~\cite{yfs:1961} and discussed in many
works implementing the YFS exponentiation;
see for example Refs.~\cite{yfs2:1990,bhlumi2:1992}.
The uncertainty of the results due to the reduction/extrapolation procedure
in the \Order{\alpha^n} YFS-exponentiated calculation is always
at least of \Order{\alpha^{n+1}};
it does not influence or spoil the \Order{\alpha^n} perturbative ``exact'' contributions
coming from the \Order{\alpha^n} Feynman-diagram calculations.
On the contrary, a reasonable choice of the reduction/extrapolation procedure
may improve significantly the total precision, because the YFS exponentiation 
is then able to sum up efficiently higher-order effects beyond the ``exact'' 
\Order{\alpha^n},
as was often seen~\cite{third-order:1991,bhlumi4:1996,bhwide:1997,CEEX:2000}.
The above remark is also valid for the special 
``multichannel'' variant of the reduction/extrapolation procedure
described in the following.

\subsection{Multichannel YFS Reduction/Extrapolation}
\label{sec:multi-reduction}

It is useful sometimes to split the $\bbeta$'s into several components and
introduce a different extrapolation/reduction procedure for each component.
This is perfectly within the limits of our freedom.
For instance in {\tt YFS3}, in the centre of mass of the $Z$-boson, neither 
the initial beams nor the final fermion pairs have the momenta back-to-back.
There are four possible definitions of the scattering angles in this frame.
One defines $\bbeta_0$ as a sum of four differential Born-level distributions, 
each for a different scattering angle. 
They might be summed up with the same weight 1/4, 
but, in fact, also a more sophisticated mixture is implemented~\cite{KKMC:2000}.
A similar multichannel extrapolation procedure is used in 
{\tt BHLUMI}~\cite{bhlumi2:1992,bhlumi4:1996}.
In the present work, we employ another kind of multichannel extrapolation, 
introduced for the first time in the unpublished program 
{\tt BHWIDE}~\cite{bhwide:1997}.

Let us elaborate on that method.
Our exponentiation model is based on the real soft photon factor $\tilde{S}$
and the corresponding virtual photon function $\Re B$ to which photons 
emitted from the beams and from the $W$'s contribute coherently.
We cannot, therefore, say whether a given photon is emitted from the initial
state or from the $W$'s.
We may, however, say something about it in a probabilistic way.
If we split $\tilde{S}$ as follows:
\begin{equation}
\begin{aligned}
\ &\tilde{S}(k) = -\frac{\alpha}{4\pi^2}
 \left( \frac{p_1}{kp_1} - \frac{p_2}{kp_2} 
      - \frac{Q_1}{kQ_1} + \frac{Q_2}{kQ_2} \right)^2
   = \tilde{S}_I(k) +\tilde{S}_W(k)+\tilde{S}_{int}(k),\\
\ &\tilde{S}_I(k) =\left( \frac{p_1}{kp_1} - \frac{p_2}{kp_2} \right)^2,\quad
 \tilde{S}_W(k) =\left( \frac{Q_1}{kQ_1} - \frac{Q_2}{kQ_2} \right)^2\quad
\end{aligned}
\end{equation}
we may define for each photon a probability
\begin{equation}
p_I(k)=\frac{\tilde{S}_I(k)}{\tilde{S}_I(k)+\tilde{S}_W(k)}
\end{equation}
that it was emitted from the beams, and the probability $p_W(k)=1-p_I(k)$
that it was emitted from the $W$'s.
The above probabilities are a kind of LL concept, which has to be used 
with care.
It may help to sum up higher-order QED corrections in the LL approximation.
In particular, it may allow us to construct a multichannel type 
of reduction/extrapolation
in which we may say for the spectator photons whether they were emitted by 
the beams (ISR) or by the $W$ bosons (WSR).
This is done as follows:
for each phase-space point we split the differential distributions into $2^n$
components, corresponding to $2^n$ possible associations of $n$ photons to 
the ISR or the WSR, the weight being the product $P(k)$ 
of the corresponding $p_I(k_i)$ or $p_W(k_i)$.
This product is exactly the partition weight $p_{\wp}$ 
in Eqs.~(\ref{Master}) and (\ref{probpar}).
For each component we may apply a separate reduction procedure which ``knows''
whether a given spectator photon belongs to the ISR or the WSR.
The above procedure is, in practice, simpler than is said above,
because in the MC realization we start from the {\tt YFS3} differential 
distribution, in which $\tilde{S}_{int}(k)$ is neglected, 
and the sum over the associations of photons to the ISR or the WSR is ``randomized''. 
In other words, we generate only one of the $2^n$ 
possible associations and therefore, for a given event, 
we work with only one component (association),
for which we may immediately employ its ``native'' reduction procedure --
that is we may use the information on the photon associations directly from 
the {\tt YFS3} MC generator;
see also the next section for an additional discussion.
Let us stress again that this powerful solution should be used with care, 
especially for the non-spectator photons in the $\bbeta$ functions.

\section{Monte Carlo Algorithm}
\label{Algor}

In constructing the MC algorithm for the cross-section calculation
and the event generation, we start from the master formula of 
Eq.~(\ref{Master}) and perform step-by-step simplifications in the
differential distribution, compensating them with appropriate
weights. 
We do this until we reach a simple enough differential distribution
to be generated using the standard MC techniques.
The MC algorithm of \yfsww\ for the Born-like $WW$ production and
decay is identical to the one used in \koralw\ for the so-called
CC03 option~\cite{koralw:1996a}, while the algorithm for the multiphoton
radiation in the 
$e^+e^- \rightarrow W^+W^-$ 
process is based on the
one implemented in the program {\tt YFS3} of {\tt KORALZ}~\cite{koralz:1994}
for the process 
$e^+e^- \rightarrow  f\bar{f}$.
Since $W$'s are much heavier than
light fermions and, in general, have different invariant masses, we had to
make a few modifications in the latter algorithm. The most important
of them are: 
\begin{itemize}
\item
  generalization of the photon radiation formulae to allow
  for large and different masses of radiating particles%
  \footnote{This extension is also included in the {\tt YFS3} 
    of the ${\cal KK}$ MC program~\cite{KKMC:2000}.},
\item
  inclusion of interferences between the radiation from the initial state and 
  that from the $WW$ state, both in the YFS IR functions $\tilde{S}$ and $Y$ and in 
  the IR-finite residuals $\bbeta$. 
\end{itemize}
The algorithm of {\tt YFS3}, although used for a decade as part of other 
programs, was not documented until recently. 
Its first full description can be found in two recent works 
\cite{KKMC:2000,CEEX:2000}.
We therefore do not describe it here in detail, but 
refer the interested reader to the above two papers.
We explain below only the main steps in the \yfsww\ algorithm that
lead to the algorithms of {\tt YFS3} and \koralw, and then comment on
the YFS reduction/extrapolation procedures.

\subsection{Main Steps in the Construction of the MC Weight} 

The MC algorithm of \yfsww\ is based on the importance-sampling method.
In constructing it, we start from the master formula~(\ref{Master})
and make a series of simplifications compensating them with appropriate
weights until we reach a distribution that is simple enough to be generated
using the basic MC methods (see e.g.~Refs.~\cite{jadach:1985,mcguide:1999}),
as we indicated above. 
Here, we describe only the main simplifications in Eq.~(\ref{Master})
that lead to the formula that can be generated with the help of 
the MC algorithms used in the programs {\tt YFS3} and \koralw.   
They are:
\begin{enumerate}
\item {\em $\bbeta$ functions:}\\
  For the aggregate of the $\bbeta$ functions we do the following replacement:
  \begin{equation}
    {\cal B}\left(\{p,Q,q,k\}^{\Rpar}\right) \longrightarrow 
    b_0\left(\{p,Q,q\}^{\Rpar}\right), 
    \label{step1}
  \end{equation}
  which is compensated by the ``model'' weight
  \begin{equation}
    w_{\beta}^{\wp} = \frac{{\cal B}\left(\{p,Q,q,k\}^{\Rpar}\right)}
                           {b_0\left(\{p,Q,q\}^{\Rpar}\right)},
    \label{w-beta}
  \end{equation}
  where $b_0=\bbeta_0^{(0)}$ is the Born-level differential cross section for
  the $W$-pair production and decay process as given in Eq.~(4) 
  of Ref.~\cite{koralw:1996a}.
  To evaluate the model weight $w_{\beta}^{\wp}$, we use the 
  partition-dependent
  reduction procedures described in Subsection~\ref{betas}.

\item {\em YFS form factor:}\\
  We simplify this with the replacement:
  \begin{equation}
    Y'(p_1,p_2,Q_1,Q_2;k_{\epsilon}) \longrightarrow 
    Y_I(p_1,p_2;\epsilon) + Y_W(Q_1,Q_2;k_{\epsilon}), 
    \label{step2}
  \end{equation}
  which is compensated by the weight
  \begin{equation}
    w_{Y} = \exp[Y' - Y_I - Y_W],
    \label{w-YFSfmf}
  \end{equation}
  where $Y_I$ is given in Eq.~(\ref{YS-ISR}) and $Y_W$ is the
  YFS IR function corresponding to the photon radiation from the $W^-W^+$ 
  electric dipole:
  \begin{equation}
    Y_W(Q_1,Q_2;k_{\epsilon}) = 2\alpha\Re B(Q_1,Q_2) 
    + 2\alpha\tilde{B}(Q_1,Q_2,k_{\epsilon}).
    \label{WWfmf}
  \end{equation}
  The explicit formulae for the above $\Re B$ and $\tilde{B}$ are 
  given in Ref.~\cite{yfsww2:1996}; 
  we do not repeat them here because they are rather lengthy. 
  This simplification corresponds to neglecting the ISR--WSR interference
  terms in the YFS form factor.
  
\item {\em $\tilde{S}$ factors:}\\
  In these we also neglect the ISR--WSR interferences:
  \begin{equation}
    \tilde{S}(p_1,p_2,Q_1,Q_2,k) \longrightarrow 
    \tilde{S}_I(p_1,p_2,k) + \tilde{S}_W(Q_1,Q_2,k), 
    \label{step3}
  \end{equation}
  and compensate this by the weight
  \begin{equation}
    w_{\tilde{S}} = \prod_{i=1}^n 
    \frac{\tilde{S}(p_1,p_2,Q_1,Q_2,k)}
    {\tilde{S}_I(p_1,p_2,k) + \tilde{S}_W(Q_1,Q_2,k)},
    \label{w-Sfact}
  \end{equation}
  where $\tilde{S}_I$ and $\tilde{S}_W$ are given in Eq.~(\ref{sisw}).
\end{enumerate}

After these simplifications, we obtain from Eq.~(\ref{Master})
the formula for the first-level, ``crude'' cross section:
\begin{equation}
\begin{aligned}
\sigma_{\rm Crude} =
&
\sum_{n=0}^\infty \frac{1}{n!}
\lint ds_1 ds_2 
\frac{d^3 Q_1}{Q_1^0} \frac{d^3 Q_2}{Q_2^0} 
\prod_{l=1}^4 \frac{d^3 q_l}{q_l^0} \;
\left[\prod_{i=1}^n  \frac{d^3 k_i}{k^0_i}
                     \left\{\tilde{S}_I(k_i) 
                         + \tilde{S}_W(k_i)\right\}\,
                     \theta(k_i^0 - k_{\epsilon}) 
\right]
\\&
\delta^{(4)}\left(p_1 + p_2 - Q_1 - Q_2 -\sum_{i=1}^n k_i \right)
\delta^{(4)}\left(Q_1 - q_1 - q_2 \right)
\delta^{(4)}\left(Q_2 - q_3 - q_4 \right)
\\&
e^{Y_I + Y_W}
\sum\limits_{\wp\in\{I,W\}^n} p_{\wp}\, b_0\left(\{p,Q,q\}^{\Rpar}\right).
\end{aligned}
\label{xcru1}
\end{equation}
The above distribution can be generated using the MC algorithms of
{\tt YFS3} and \koralw. 
To make this more transparent, we shall write this formula in a slightly
different but equivalent form.

The product of the $\tilde{S}$-factors in the square brackets of 
Eq.~(\ref{xcru1}) together with the sum over the photon partition 
can be written as 
\begin{equation}
\begin{aligned}
\ &
\prod_{i=1}^n  \frac{d^3 k_i}{k^0_i}
                     \left\{\tilde{S}_I(k_i) 
                         + \tilde{S}_W(k_i)\right\}\,
                     \theta(k_i^0 - k_{\epsilon})  
\sum\limits_{\wp\in\{I,W\}^n} p_{\wp}\, b_0\left(\{p,Q,q\}^{\Rpar}\right) = 
\\&
\sum_{\wp \in \{I,W\}^n}\;
  \prod_{\wp_i = I} \frac{d^3 k_i}{k^0_i}\tilde{S}_I(k_i)
  \theta(k_i^0 - k_{\epsilon})
  \prod_{\wp_j = W} \frac{d^3 k_j}{k^0_j}\tilde{S}_W(k_j)
  \theta(k_j^0 - k_{\epsilon})\,
  b_0\left(\{p,Q,q\}^{\Rpar}\right),
\end{aligned}
\label{S-part}
\end{equation}
where we have substituted Eq.~(\ref{probpar}) for the partition weight $p_{\wp}$.

After some algebra, and exploiting the Bose--Einstein symmetry for photons,
the formula of Eq.~(\ref{xcru1}) can be written as
\begin{equation}
\begin{aligned}
\sigma_{\rm Crude} =
&
\sum_{n_I=0}^\infty \sum_{n_W=0}^\infty \lint 
ds'\frac{d^3 P'}{{P'}^0}\,
\delta^{(4)}\left(p_1 + p_2 -\sum_{i=1}^{n_I} k_i - P' \right)
ds_1 ds_2  
\frac{d^3 Q_1}{Q_1^0} \frac{d^3 Q_2}{Q_2^0}\,
\\&
\left[\frac{1}{n_I!} \prod_{i=1}^{n_I} 
                     \frac{d^3 k_i}{k^0_i}\tilde{S}_I(k_i)\,
                     \theta(k_i^0 - k_{\epsilon}) \right]
\left[\frac{1}{n_W!} \prod_{j=1}^{n_W}  
                     \frac{d^3 k_j}{k^0_j}\tilde{S}_W(k_j)\,
                     \theta(k_j^0 - k_{\epsilon}) \right]
\\&
\delta^{(4)}\left(P' - Q_1 - Q_2 -\sum_{j=1}^{n_W} k_j \right)
\prod_{l=1}^4 \frac{d^3 q_l}{q_l^0}\,
\delta^{(4)}\left(Q_1 - q_1 - q_2 \right)
\delta^{(4)}\left(Q_2 - q_3 - q_4 \right)\,
\\&
e^{Y_I} e^{Y_W}\,
b_0\left(\{p,Q,q\}^{\Rpar}\right),
\end{aligned}
\label{xcru1a}
\end{equation}
where $s'={P'}^2$ is the CMS energy squared in the ISR-reduced (effective) 
frame, and $n_I$ and $n_W$ are the multiplicities of the ISR and WSR photons, 
respectively. 

To generate the ISR and WSR according to Eq.~(\ref{xcru1a}),
we can use the MC algorithm of {\tt YFS3} 
(see Refs.~\cite{KKMC:2000,CEEX:2000}) for multiphoton radiation in the 
fermion-pair production, where the initial-final-state interferences 
were neglected. The only modification we had to do in this algorithm
was to extend it to the case of heavy particles of unequal masses
(see Ref.~\cite{KKMC:2000}).
Generating particular values of $n_I$ and $n_W$ 
corresponds to choosing a given photon partition for $n=n_I+n_W$
in the sum of Eq.~(\ref{S-part}). This means that the sum over
partitions is ``randomized'' in the program, i.e. not evaluated directly 
but generated using the MC methods.  
For the given (generated) photon partition, we then calculate the weights 
$w_{\beta}^{\wp}$, $w_{Y}$  and $w_{\tilde{S}}$.
Such a solution can be regarded as a multibranch MC algorithm, 
see Ref.~\cite{mcguide:1999},
where each branch corresponds to a particular photon partition. 

\subsection{MC Weights and Absolute Normalization}

In the process of simplifying the fully differential distribution of 
Eq.~(\ref{Master})
\begin{equation}
\sigma_{\rm Best} = 
\sum_{n=0}^\infty \frac{1}{n!} \int d{\rm Lips}_n(\{p;q\},k_1,\ldots,k_n)
\rho^{\rm Best}_n(\{p;q\},k_1,\ldots,k_n)
\end{equation}
described in the previous section we have finished with the 
``crude differential distribution''
$\rho^{\rm Crude}$ inside the integral
\begin{equation}
\sigma_{\rm Crude} = 
\sum_{n_I=0}^\infty \sum_{n_W=0}^\infty
\frac{1}{n_I!}\frac{1}{n_W!}
\int d{\rm Lips}_n(\{p;q\},k_1,\ldots,k_n)\;
\rho^{\rm Crude}_{n_I,n_W}(\{p;q\},k_1,\dots,k_n),
\end{equation}
where $n=n_I+n_W$.
Using the above notation,
the total correcting weight is defined as follows:
\begin{equation}
\label{eq:wtmodel}
  w^{(1)}=\frac{\rho^{\rm Best}_n(\{p;q\},k_1,\ldots,k_n)}
                    {\rho^{\rm Crude}_{n_I,n_W}(\{p;q\},k_1,\dots,k_n)}
              =w_{\beta}^{\wp}\; w_{Y}\; w_{\tilde{S}}.
\end{equation}
In the actual program it comes in several versions, because for the purpose
of the technical tests, and for the discussion of the physical precision,
we need to switch off/on certain contributions in the full distribution 
$\rho^{\rm Best}_n$. 

However, in the actual MC program we do not actually generate the distribution 
$\rho^{\rm Crude}$,
but a certain, more primitive, ``primary'' distribution $\rho^{\rm Prim}$.
For its definition and full description,
and the description of how events are generated according to this distribution,
starting from the uniform random numbers, we refer the reader to the most
complete documentation of the {\tt YFS3} Monte Carlo algorithm 
in Ref.~\cite{KKMC:2000};
Refs.~\cite{koralw:1996a} and \cite{yfs2:1990} can also be helpful.
The weight correcting for the transition from the primary
to crude distribution is given by
\begin{equation}
  w^{(2)}=\frac{\rho^{\rm Crude}_{n_I,n_W}(\{p;q\},k_1,\ldots,k_n)}
                     {\rho^{\rm Prim}_{n_I,n_W}(\{p;q\},k_1,\ldots,k_n)}
\end{equation}
and is fully defined in Ref.~\cite{KKMC:2000}.
The total MC weight is of course equal to the product
\begin{equation}
  w=w^{(1)}\; w^{(2)}
\end{equation}
or 
\begin{equation}
  w=w^{\rm Model}\; w^{\rm Crude},
\label{wt-total}
\end{equation}
with 
\begin{equation}
  w^{\rm Model}=w_{\beta}^{\wp} \; b_0, \hspace{1cm}
   w^{\rm Crude}= w^{(2)}\, w_{Y}\, w_{\tilde{S}} \; b_0^{-1},
\label{wt-model}
\end{equation}
where $w^{\rm Model}$ is called in the program the ``model weight''
(this weight and its many variants are provided in 
the {\tt WtSet} array; see Section 5),
and $w^{\rm Crude}$ is called the ``crude weight'' 
(in the program it is equal to 
the product {\tt WtCrud1*WtCrud2}; see Section 5).

The integrated cross section is given by the product of the average weight
and the integrated primary cross section
\begin{equation}
\label{eq:absolute}
  \sigma = \langle w \rangle \, \sigma^{\rm Prim} = \langle w \rangle \int d\sigma^{\rm Prim}.
\end{equation}
The integrated primary cross section we also call
a ``normalization cross section'',
because it provides the absolute normalization for the
whole MC calculation.

\section{Structure of the Program}

In this section we provide the reader with a brief guide of the \yfsww\
program. We shall describe its main routines, libraries and interfaces.
We want to note here that the program, in the distribution version, is
prepared for a UNIX/Linux-type operating system that supports directories 
and the {\tt make} utility. 
The source code is distributed over a number of subdirectories, 
in order to make the structure of the program more
transparent and easier to handle. 
Also a system of the {\tt Makefile}'s is  provided, for compilation, 
execution and other auxiliary functions (e.g. clean-up). 
(Of course, this organization, which is in fact rather simple, 
can be avoided and the whole source code can be put into a single
FORTRAN file.)
In the following we give a review of all \yfsww\
subdirectories. We start from the directories that are most interesting 
from the user's point of view, then going to the ones with more technical 
contents.
In the distribution package, all these subdirectories are located in the main
\yfsww\ directory: {\tt yfsww3-1.16-export}. This directory also
comprises two important files: {\tt README} and {\tt RELEASE.NOTES},
which we recommend the user look through before using the program.
They contain some basic information about \yfsww: how to compile/link
and run the program, a brief documentation of the code, etc. 

\subsection{{\tt demo} -- Demonstration Program}

The subdirectory {\tt demo} contains a demonstration program in 
the file {\tt demo.f}.
Generally, the user is supposed to provide his/her own main program;
nevertheless, the file quoted here provides a simple example
of such a program. It has a double role: (1) as a useful template, 
and (2) as a first cross-check that the MC generator \yfsww\ 
runs correctly on a given installation.
The essential part of this program is a loop in which a series of MC 
events is generated.
It also reads the input from a disk file, but no histogramming is performed
and most of the output comes from the generator itself.  
At the end of the program, a MC-integrated cross section of \yfsww\  
is compared with the Born-level and ISR results from the semi-analytical
program \korwan.
The program is compiled/linked and executed, with the help of {\tt Makefile}, 
for the input data set given in the files {\tt demo.input}.  
The program also reads the default settings provided in the file
{\tt data\_DEFAULTS} located in the directory {\tt data\_files} -- some
of these default settings are overridden by data from {\tt demo.input}.
The demonstration program is run for unweighted events with the
external libraries \photos, \tauola\ and \jetset\ switched
on. The output, written into the disk file {\tt demo.output}, can then be
compared with the one provided in the file {\tt demo.output.linux}. 
It was obtained on a PC Intel Pentium III under
the Linux RedHat 6.1 operating system.

\subsection{{\tt data\_files} -- Default Input Data}

The subdirectory {\tt data\_files} contains the input data file 
{\tt data\_DEFAULTS},
which gives the defaults settings for \yfsww. It is {\em identical} with the one 
used in the program \koralw, so the two programs can be set-up 
from the same input data file! This considerably facilitates a combination
the results of the two programs; see Ref.~\cite{koralw:2001} for more details.
Some entries in this data file are dummy for \yfsww\ and are kept
only for compatibility with \koralw, while some of them are specific
to \yfsww\ and are treated as dummy parameters in \koralw.
They are accompanied by appropriate comments. 
This file can also be used  as a template to create the user's
own input data file, where only those entries that will have different 
settings from the default data file may be included.
The user input file should be read after the file
{\tt data\_DEFAULTS},  so that the user settings can override 
the default ones; see the example in the {\tt demo.f} file described in 
the previous subsection.
This directory comprises also some older data files kept for backward
compatibility. Their names contain, after a dot, the last two digits of
the year of their creation. 
(The latest of these files is identical with the file {\tt data\_DEFAULTS}.)

\subsection{{\tt yfsww} -- Master Unit}
\label{yfsww}

The subdirectory {\tt yfsww} contains the actual Monte Carlo event generator. 
Subprograms that can be used directly are the following:
\begin{itemize}
\item 
  {\tt YFSWW\_ReaDataX}  
  -- the subprogram used to read, from the disk file, the default 
  input data of \yfsww\ and subsequently the data of the user
  into the array {\tt xpar} at the very beginning of the use of \yfsww.
\item 
  {\tt YFSWW\_Initialize} 
  -- the subprogram that does all
  initializations of internal variables. First, it sets-up the main 
  parameters of the program (through the routine {\tt filexp}) according 
  to the values in the input data files, then it calls several
  {\em initializers} of the main internal modules of the generator,
  such as {\tt KarLud(-1,...)}, and of external packages, such as \tauola,
  etc.
  It prints out directly or indirectly all the input parameters.
\item 
  {\tt YFSWW\_Make} 
  -- the most important subprogram of \yfsww. 
  It generates single MC events.
  Functionally, it is a high-level management subprogram in the 
  event generation. 
  It invokes (through the routine {\tt yfsww3}) other routines that perform 
  specific tasks, such as:
  generation of ISR photons ({\tt KarLud}), generation of radiative
  photons from $W$'s ({\tt KarFin}), generation of the $W$'s and final $4f$ 
  four-momenta ({\tt WW\_Presam}), evaluation of the Coulomb correction 
  ({\tt WTCoul}), calculation of the matrix element ({\tt Model)}, of the 
  YFS form factor, of the interference corrections, etc.
  It builds up the total MC weight from the weights supplied by these routines. 
  This weight is then returned, in the weighted-event mode,
  as the main event weight,
  or used in the rejection loop (in the routine {\tt yfsww3}), 
  in the unweighted-event mode, for constructing the weight $=1$ event.
  It also calls \photos, which generates photon radiation in the $W$ decays,
  \tauola\ for decaying of the final-state $\tau$'s (if they appear),
  and \jetset\ to perform fragmentation/hadronization of the final-state quarks. 
  It keeps track, through special monitoring routines, of all the information
  necessary for the final results from the MC event generation.
\item 
  {\tt YFSWW\_Finalize} 
  does all final bookkeeping, including the calculation of the integrated 
  (total) cross section. It prints a summary output for the whole sample 
  of generated MC events.
\end{itemize}
The first subprogram is located in the file {\tt readata.f}, 
while the following three are in the file {\tt yfsww3.f}.

This directory contains also a collection of utility routines,
which can be used, for instance, to access information on particle flavours 
and four-momenta, on event weights, etc. 
They are located in the file {\tt ww\_get.f} 
(see the comments in this file for more details).

\subsection{{\tt model} -- Matrix Elements}
\label{model}

The routines for the ``model'' weights calculations are located in the
subdirectory {\tt model}. 
The master routine here is the subroutine {\tt Model}
(in the file {\tt model.f}); by calling some other routines, this
evaluates the ``model'' weights $w^{\rm Model}$ as defined in 
Eq.~(\ref{wt-model}). 

\begin{enumerate}
\item
  For {\bf the Born-level matrix element} we use
  the same routines as in the program \koralw\ -- they are collected
  in the file {\tt born.f}. The same again are used for the ISR corrections, located
  in the file {\tt betas.f}. 
\item
  {\bf The \oal\ virtual and real soft photon corrections} 
  are evaluated with the help of a special interface 
  -- the routine {\tt VirSof} --  in the file {\tt virsof.f}.
  In the first call, this routine makes all the necessary initializations,
  and then calculates the EW corrections by calling appropriate routines 
  from the EW library for each generated event. 
  The pretabulated (or ``fast'') version of the EW correction is invoked 
  through the routine {\tt DnlFast} (in the file {\tt dnlfast.f}). 
  This routine makes a pretabulation of the EW corrections in 
  the initialization stage of the program and then, during the event
  generation, it performs the linear interpolation of the pretabulated values 
  to calculate the EW corrections for each event as described in 
  Section~\ref{pretab}.
\item 
  {\bf The \oal\ hard photon matrix element} in the $WW$ production, which
  contributes to the function $\bbeta_1$, see Eq.~(\ref{beta1}),
  is calculated with the help of the routine {\tt eewwg}, located
  in the file {\tt eewwg.f}. This routines evaluates appropriate
  spin amplitudes for the process 
  $e^+e^-\rightarrow 
  W^+W^-\gamma$
  according to the formulas of Ref.~\cite{kolodziej:1991},
  which are then combined coherently with the spin amplitudes
  for $W$ decays.
\end{enumerate}
In order to calculate all these contributions to the $\bbeta$ functions,
the subroutine {\tt Model} calls the routines that perform appropriate 
``reduction'' procedures as described in Section~2. 
Except for calculating the contributions to the YFS exponentiated
cross section, this routine provides also the weights corresponding
to the strict \Order{\alpha^0} and \Order{\alpha^1} calculations. 
In the end, the subroutine {\tt Model} fills the array of weights with 
the weights corresponding to:
(i)  several variants of the SM models (for example different orders of QED ISR) and
(ii) various components in the differential cross section for a given
    variant of the SM (for example contributions from various $\bbeta$'s).
The subroutine {\tt Model} returns the best ``model'' weight.

The routine {\tt WTCoul} -- in the file {\tt coulco.f} -- for
the Coulomb (both ``standard'' and ``screened'') correction 
calculation is also located in this directory. 

\subsection{{\tt ewc} -- Electroweak Library}

The subdirectory {\tt ewc} contains the routines for the calculation of the  
electroweak (virtual plus soft photon) corrections in the on shell 
$WW$ production process according to Refs.~\cite{fleischer:1989,fleischer:1995}.
For the evaluation of one-loop integrals they use the package 
{\tt FF} of Ref.~\cite{ff:1990}, which is located in a separate 
directory (see below).

\subsection{{\tt interfs} -- Interfaces to External Libraries}

Interfaces to some external libraries are collected in the directory 
{\tt interfs}.
The most important routines are:
\begin{itemize}
\item
  {\tt inietc} -- in the file {\tt tauola\_photos\_ini.f} -- sets up 
  some parameters of \photos\ and \tauola.
\item
  {\tt tohep} -- in the file {\tt hepface.f} -- fills in the {\tt /HEPEVT/} 
  {\tt COMMON} block, calls \photos\ for photon radiation in $W$ decays and 
  \tauola\ for $\tau$ decays.
\item
  {\tt tohad} -- in the file {\tt lundface.f} -- calls \jetset\ 
  for the quark fragmentation/hadronization.
\end{itemize}
Note that the {\tt COMMON} block {\tt /HEPEVT/} is expected to contain
single-precision ({\tt REAL*4}) variables.

\subsection{{\tt wdeclib} -- External Libraries for $W$ Decay}

The subdirectory {\tt wdeclib} provides the following programs: 
\begin{itemize}
\item
  \photos~\cite{photos:1994} for photon radiation in the $W$ decays 
  (up to two photons),
\item
  \tauola~\cite{tauola:1993} for $\tau$ decays with radiative corrections,
\item
  \jetset\ version {\tt 7.4}~\cite{jetset:1987} 
  for quark fragmentation/hadronization,
\end{itemize}
located in the files: {\tt photos.f}, {\tt tauola.f} and {\tt jetset74.f},
respectively.

In {\tt YFSWW3~1.16} we use the recently modified \photos, 
where also the photon radiation from quarks can be activated with 
the help of a special input parameter switch; 
see the tables with the input parameters in Appendix~A.
Note that in the distribution version of \tauola\ the parameters 
in the $\tau$-decay modes are not adjusted to the recent experimental data.
We recommend the user to replace this version of \tauola\ 
with the one accepted within his/her own collaboration.   
The technical update described in Refs.~\cite{golonka:2000,pierzchala:2001}
will resolve this inconvenience in the future releases of the program.

\subsection{{\tt glib} -- Histogramming and Numerical Library}

A handy FORTRAN histogramming package, {\tt GlibK}~\cite{glibk:1995}, 
is provided in the subdirectory {\tt glib} (the file {\tt glibk.f}). 
It is used by \yfsww\ both for hard-coded internal 
bookkeeping and for some optional ``external''  tests. 
The package is similar in its usage to the classic {\tt HBOOK} 
of CERNLIB.
This subdirectory also contains the useful numerical package {\tt yfslib.f}
with the routines for weight monitoring, random-number generation, numerical 
integration, one- and two-dimensional adaptive MC sampling, Lorentz 
transformations, etc.

\subsection{{\tt semian} -- Semi-Analytical Program \korwan}

The subdirectory {\tt semian} contains the package \korwan\ for the 
semi-analytical 
calculations of the cross section for the off-shell $WW$ production and decay
at the Born level and with the ISR corrections (using the structure-function
formalism). Also the Coulomb effect is included in \korwan\ in both versions:
the ``standard'' and the ``screened'' ones.
The package is described in detail 
in Refs.~\cite{koralw:1996a,koralw:1996b,koralw:1999}.
It can be used for a quick consistency check of the \yfsww\ results.

\subsection{{\tt ff} -- {\tt FF} Package}

In the subdirectory {\tt ff} is located the {\tt FF} package~\cite{ff:1990} 
for the evaluation of one-loop integrals needed to calculate 
the \oal\ EW virtual corrections. 

\subsection{{\tt dok} -- Related Papers}

In the subdirectory {\tt dok}, we include some of our papers related 
to \yfsww. They are in the form of {\tt gzip}'ed
PostScript files.

\subsection{{\tt rewt} -- Reweighting Tools}
\label{rewt}

The subdirectory {\tt rewt} contains routines that are not needed for 
the standard MC event generation but for various event reweighting purposes.
We provide here the tools for two kinds of reweighting: the 
reweighting (1) of the \koralw\ generated events to correct for \oal\ NL effects
in $WW$ production, and (2) of the \yfsww\ generated events
to take into account effects due to some input parameters (typicaly $W$-mass) changes.
All these tools are described in detail in the following; see also
the file {\tt README} provided in this  subdirectory. 

\subsubsection{Reweighting \koralw\ Events by \yfsww}

The main tool for this kind of reweighting is the function 
{\tt YFSWW\_WtNL}, located in the file {\tt rewt\_K.f}.
For the events, in terms of flavours and four-momenta of the final-state 
fermions and the ISR photons, provided through its parameters,
this function calculates and returns the corresponding correction
weight from \yfsww. This weight is a ratio of the fully differential 
distribution.
In practice, we calculate the ratio of the MC weights 
corresponding to two distributions -- this is more convenient.
We can do it because the common crude distribution entering both 
weights cancels out,
for a given number of real photons of Eq.~(\ref{eq:rho}), 
to the reference distribution in the ``ISR approximation'' 
defined in Eq.~(\ref{eq:rhoISR}),
which is the same as in \koralw\ in the CC03 mode 
(Eq.~(4) of Ref.~\cite{koralw:1999}).
It is defined in Eq.~(\ref{eq:deltaNL}).
In the reweighting mode of \yfsww, we implement an approximate version
of this weight:
\begin{equation}
  w^{-}_{\rm NL} \equiv 
  \frac{\rho_n^{{\rm Best}-}(\{p,Q,q\}^{\Rcal_I},k_1,\ldots,k_{n})}
       {\rho_n^{R}(\{p,Q,q\}^{\Rcal_I}, k_1,\ldots,k_{n})}
  = 1 + \delta^R_{\rm NL-},
  \label{wnl}
\end{equation}
where $\rho^{{\rm Best}-}_{n}$ is a variant of $\rho^{{\rm Best}}_n$
in which the sum over the photon partitions is reduced to only
one term, in which all photons are associated to the ISR with the 
partition probability $p_{\wp}=1$, and $\Rcal_I$ means the respective
reduction procedure.  
Unfortunately, we loose in this way a (small) part of the genuine \oal\ 
corrections.
The full \oal\ correction can be included only if \yfsww\
is run in the event-generation mode, and the correction weight 
for reweighting (accounting for the $4f$ background) is supplied from \koralw. 
(This leads, however, to large MC weight fluctuations, see below.)
We have checked, however, that the bias introduced in the LEP2 energy range
by this approximation is within $0.1\%$, 
both for the total cross section and for distributions. 
This can be explained by the fact that the $W$'s, as heavy and slowly 
moving particles, do not radiate much at these energies. 
Of course, the main weight of \yfsww\ based on $\rho^{{\rm Best}}_n$ does
not have this bias.
In this way, by assigning this correction weight to the \koralw\
events, one can include, in a simple way, the \oal\ NL correction 
$\delta^R_{\rm NL-}$ into the \koralw\ results. How to do it in practice, 
we explain in detail in Ref.~\cite{koralw:2001}.

\subsubsection{Reweighting \yfsww\ Events by \yfsww}

This kind of reweighting can be useful in some data analyzes,
e.g. when one has accumulated a large sample of simulated events for a given
input parameter set-up and one wants to assess the effects of changing
some of the input parameters, but without having
to repeat the full simulation with the new input parameters.
The tools for such a reweighting are collected
in the file {\tt rewt\_Y.f}. We provide two methods: ``exact'' and 
``approximate''.
In the exact method the partition vector
for photons is stored along with their four-momenta, while in the approximate
method the partition information is not recorded.

\vspace{3mm}
\noindent
{\bf (a) The exact method:}

\vspace{1mm}
\noindent
The tool for the exact method is {\tt SUBROUTINE YFSWW\_WME(WtME)},
where {\tt WtME} corresponds to the weight $w^{\rm Model}$ of 
Eq.~(\ref{wt-model}). 
It requires storing on the disk/tape the entire content of four {\tt COMMON} 
blocks: {\tt /MOMINI/, /MOMFIN/, /MOMDEC/} and {\tt /DeChan/}
during the event-generation run of \yfsww. We assume that the appropriate
tools for writing the above information on the disk/tape as well as for 
reading it in the reweighting run of the program are provided by the user 
of the code (see the source code for more details).

\vspace{1mm}
\noindent
\underline{How to use it:}
\begin{enumerate}
\item
 {\em The primary-run generating/storing events:} \\
  $\bullet$ Run \yfsww\ and store on the disk/tape the whole content of 
  the {\tt COMMON} blocks:  
  \mbox{\hspace{3mm}} {\tt /MOMINI/, /MOMFIN/, /MOMDEC/, /DeChan/}.\\
  \mbox{\hspace{3mm}} This is done for some input parameters set-up {\sf (A)}. 
\item
 {\em The secondary-run reweighting events:}\\
  $\bullet$ Initialize the program with the input parameters set-up {\sf (A)} 
            used for the event \mbox{\hspace{1mm}}
            \mbox{\hspace{3mm}} generation. \\
    $\bullet$ Instead of generating events, call {\tt YFSWW\_WME(WtME\_A)} 
    for all stored events and \mbox{\hspace{1mm}} \mbox{\hspace{3mm}} 
    save the computed 
    matrix element weight {\tt WtME\_A} (in some arbitrary units).\\
   $\bullet$ Initialize the program with a different input parameters 
    set-up {\sf (B)}.\\
   $\bullet$ Calculate the weight {\tt WtME\_B} and save as above.\\
   $\bullet$ For each event, calculate the ratio {\tt Wt = WtME\_B/WtME\_A} 
     and use it as the weight \mbox{\hspace{3mm}} 
     for reweighting the original sample of events.\\
   $\bullet$ This procedure can be repeated for any new input parameters 
    set-up {\sf (B)}.
\end{enumerate} 
\underline{Example of running the program in the reweighting mode:}
{\small
\begin{alltt}
 CALL YFSWW\_ReaDataX('./data\_DEFAULTS',1,10000,xpar)  ! reading general defaults
 CALL YFSWW\_ReaDataX('./user.input'   ,0,10000,xpar)  ! reading user input
 CALL YFSWW\_Initialize(xpar)                          ! initialize generator 
 NevTot = 1000            ! number of events to be reweighted
 DO Iev = 1, NevTot
    CALL YFSWW\_WME(WtME)  ! calculate matrix element weight for stored events
 ENDDO    
\end{alltt} 
}
\noindent
For more information about running \yfsww, see Section~\ref{Howtouse}.

\vspace{3mm}
\noindent
{\bf (b) The approximate (simpler) version:}

\vspace{1mm}
\noindent
In the case when one does not have all the information needed for the exact
(recommended) method of reweighting (because of the lack of disk/tape space
or some other reasons), one can use 
\begin{alltt}       
     WtME = YFSWW_WME\_Smpl(Iflav,pf1,pf2,pf3,pf4,Phot,Nphot)
\end{alltt}
instead of {\tt CALL YFSWW\_WME(WtME)}. 
In this case one needs to store for each event only the flavours and 
four-momenta of the final-state fermions 
and the number and the four-momenta of all photons. 
{\em Important:} They should come from \yfsww\ itself, not from \photos\
after the FSR. 
These four-momenta and flavours are available from two {\tt COMMON} blocks:
\begin{alltt}
     COMMON / DECAYS / Iflav(4),amdec(4)                              
     COMMON / MOMDEC / pf1(4),pf2(4),pf3(4),pf4(4),Phot(100,4),Nphot 
\end{alltt}
(the fermion mass array {\tt amdec(4)} does not need to be stored). 
They can be alternatively accessed with the help of two getter-routines:
\begin{alltt}
     SUBROUTINE YFSWW_Get4f(flav,p1,p2,p3,p4)
     SUBROUTINE YFSWW_GetPhotAll(NphAll,PhoAll)
\end{alltt}
located in the file {\tt ww\_get.f} in the subdirectory {\tt yfsww}.
The first one provides the flavours and four-momenta of the final-state 
4-fermions, while the second one provides the number and 
four-momenta of the radiative photons (excluding the ones from \photos).

The {\tt FUNCTION YFSWW\_WME\_Smpl} fills the appropriate {\tt COMMON} blocks 
and calculates (returns) the matrix element weight (corresponding to 
{\tt WtME} in the previous method). 
This method is {\em approximate} because it does not use the information 
on the photon partitions but assigns all the photons to the ISR in performing
the partition-dependent reduction procedure 
(described in Section~\ref{MastFor}). 
Our tests show, however, that the differences between this and the exact
method are numerically negligible, e.g. $<0.01\%$ for the total cross 
section at $\sqrt{s} = 200$~GeV.

\subsubsection{Reweighting \yfsww\ Events by \koralw}

Similar reweighting tools are provided in the new version of 
\koralw~\cite{koralw:2001}. 
They can be used for reweighting events generated by \yfsww\
with the weight provided by \koralw\ in order to correct for 
the missing $4f$-background contribution to the signal $WW$ process.  
The relevant reweighting routines in \koralw\ 
require as an input the events from \yfsww, which are
generated according to the ISR-type reference differential distribution.
Such events are provided by \yfsww\ in the standard generation mode,
with the help of the getter-routine {\tt YFSWW\_GetEvtISR},
located in the file {\tt ww\_get.f} of the directory {\tt yfsww}.
The correction weight provided by \koralw\ can be used to reweight the
constant-weight (unweighted) events from \yfsww\ or to correct the \yfsww\
weights in the variable-weight (weighted-event) mode;
see Ref.~\cite{koralw:2001} for more details. 

\section{How to use the Program}
\label{Howtouse}
In this section we provide a short guide of how to use the current version
of \yfsww, and we describe the input parameters of the program and its output.

\subsection{Principal Entries of \yfsww}

The principal entries of the \yfsww\ package, which the user has to call in
his/her application in order to generate a series of MC events, were
already listed and described briefly in Subsection~\ref{yfsww}.
Here we shall add more information on their functionality.
The calling sequence constituting a typical Monte Carlo run
will look as follows:
{\small
\begin{alltt}
 CALL YFSWW\_ReaDataX('./data\_DEFAULTS',1,10000,xpar)  ! reading general defaults
 CALL YFSWW\_ReaDataX('./user.input'   ,0,10000,xpar)  ! reading user input
 CALL YFSWW\_Initialize(xpar)                          ! initialize generator 
 DO loop = 1, 10000                                   ! loop over MC events
   CALL YFSWW\_Make                                    ! generate single MC event
 ENDDO    
 CALL YFSWW\_Finalize                                  ! final bookkeeping, print
 CALL YFSWW\_GetXSecMC(XSecMC,XErrMC)                  ! get total cross section
\end{alltt} 
}
\noindent
In the first call of {\tt YFSWW\_ReaDataX}, default data are read into 
the {\tt REAL*8} array {\tt xpar(10000)}.
The \yfsww\ has almost no data hidden in the source code
(this is not true for \tauola\ and \jetset).
The file {\tt data\_DEFAULTS}, which is read first,
is placed in the distribution subdirectory {\tt data\_files}.
This file provides necessary initial default values of {\em all} input parameters.
The user should {\em never modify it}. It can be copied to a local directory
or, better, a symbolic link should be created to the original 
{\tt data\_files/data\_DEFAULTS}. 
This file is quite sizeable and the user is usually interested
only in changing some subset of these data.
In the second call of {\tt YFSWW\_ReaDataX}, the user can overwrite the default
data with his/her own smaller set of input data, which are placed in the
{\tt user.input} file.
For example, the simplest input data, defining only the CMS energy,
would look like this:
\begin{alltt}
BeginX
*<ia><----data-----><-------------------comments------------->
    1          190d0 CMSEne  =CMS total energy [GeV]
EndX
\end{alltt}
As we see, data cards begin with the keyword {\tt BeginX} and end with  
the keyword {\tt EndX}.
The comment lines are allowed -- they begin with {\tt *} in the first column.
The data themselves are in a fixed format, with the index {\tt i} of the array 
{\tt xpar(i)} followed by the data value and a trailing comment.
The example of the input data set for the demonstration program 
{\tt demo.f} in the subdirectory {\tt demo} provides a useful template 
for typical user's data.
The complete set of data in {\tt data\_DEFAULTS} is described in
detail in Tables~\ref{tab:yfsww-input1}--\ref{tab:yfsww-input5}, see Appendix A.
Obviously, the user is interested in manipulating only some of them
and will retain the default values in most of the cases.

The {\tt YFSWW\_Initialize} is invoked to initialize the generator.
It reads the input data from the array {\tt xpar}, prints them out 
and sends them down to the various modules and auxiliary libraries.
The programs have to be called strictly in the same order
as in the above example.
At this point one is ready to generate the series of MC events.
The generation of a single event is done with the help of {\tt YFSWW\_Make}.
After the generation loop is completed, we may invoke {\tt YFSWW\_Finalize},
which does the final bookkeeping, prints out various pieces of information 
on the MC run, and calculates the total MC integrated cross section (in pb).
In order to obtain this cross section the user may call the routine
{\tt YFSWW\_GetXSecMC(XSecMC,XErrMC)}.

\subsection{Input/Output Parameters}
\label{sec:inp-outp}

As we explained in the previous section, the input
parameters enter through the {\tt xpar} array, being a parameter of 
the routine {\tt YFSWW\_Initialize}.
Their meaning is explained in Tables~\ref{tab:yfsww-input1}--\ref{tab:yfsww-input5}
of Appendix A.

The principal output of \yfsww\ is the Monte Carlo {\em event},
which is just a list of final-state four-momenta (in GeV) 
and flavours, encoded in the standard {\tt /HEPEVT/} event record.
In the present version, we still provide the {\tt REAL*4} version of 
{\tt /HEPEVT/} of dimension {\tt 2000}.
If the user is interested in the parton momenta before hadronization, then,
in addition to {\tt /HEPEVT/}, they are available
(see also Table~\ref{tab:yfsww-event}) through
{\small
\begin{alltt}
 CALL YFSWW\_Get4f(p1,p2,p3,p4)         ! get final 4f flavours and 4-momenta  
 CALL YFSWW\_GetBeams(q1,q2)            ! get beam 4-momenta
 CALL YFSWW\_GetPhotAll(NphAll,PhoAll)  ! get photon multiplicity and 4-momenta
\end{alltt} 
}
\noindent
Alternatively, all the four-momenta are available from
the internal {\tt COMMON} blocks {\tt /MOMWWP/} and {\tt /MOMDEC/}, see Tables 
\ref{tab:yfsww-momwwp} and \ref{tab:yfsww-momdec} for details. 

In the case of a MC run with weighted events, the user is provided with
the main weight (see also Table~\ref{tab:yfsww-getwgt}) through
{\small
\begin{alltt}
 CALL YFSWWW\_GetWtMain(WtMain)              ! get main Monte Carlo weight
\end{alltt} 
}
\noindent
Of course, for unweighted events {\tt WtMain=1}.
For special purposes the user may also be interested in auxiliary weights,
which are provided (see also Table~\ref{tab:yfsww-getwgt})
with the help of
{\small
\begin{alltt}
 CALL YFSWW\_GetWtAll(WtMain,WtCrud,WtSetAll)  ! get all Monte Carlo weights
\end{alltt} 
}\noindent
where {\tt REAL*8 WtSetALL(100)} is an array of the weights described
in Table \ref{tab:yfsww-weights}.
The definition of the main weight {\tt WtMain} depends 
on the input switches {\tt KeyCor} and {\tt KeyLPA},
see Table~\ref{tab:yfsww-weights}; for instance, the best model 
for {\tt KeyCor = 5} and {\tt KeyLPA = 0} corresponding to Eq.~(\ref{Master})
is defined as follows:
{\small\begin{alltt}
 WtMod = Wtcru1*Wtcru2*(WtSetAll(41)-WtSetAll(2)+WtSetAll(4)).
\end{alltt}}\noindent
Alternatively, the auxiliary weights {\tt WtSet} defined 
in Eq.~(\ref{wt-model}) are available from
the {\tt COMMON /WGTALL/}, see Table \ref{tab:yfsww-weights}.
The total auxiliary weight should be defined as: 
{{\small\begin{alltt}
 WtAuxEWC(i) = WtCru1*WtCru2*WtSetAll(i)
\end{alltt}}\noindent
for \oal\ EW corrections, and 
{{\small\begin{alltt}
 WtAuxISR(i) = WtCru1*WtCru2*WtSetAll(i)
\end{alltt}}\noindent
for the ISR corrections.
The corresponding integrated cross section is simply obtained by
multiplying the average of the total weight by the {\em normalization} 
cross section of Eq.~(\ref{eq:absolute}) 
(or the ``primary cross section'' in the terminology 
of Ref.~\cite{KKMC:2000}), which is provided through
{\small
\begin{alltt}
 CALL YFSWW\_GetXSecNR(XSecNR,XErrNR)       ! get normalization x-section
\end{alltt} 
}\noindent
Note that we expect the user to exploit the auxiliary weights
only in the variable-weight operation mode.
It is, however, not impossible to use them also for the constant-weight events.
In such a case we recommend the user to contact the authors for more 
instructions on how to do it correctly.

The complete description of the post-generation
output parameters from {\tt YFSWW\_Finalize} is collected in 
Table~\ref{tab:yfsww-final}.

\subsection{Printouts of the Program}
\label{PRINTOUTS}

In this section we describe a printout of the demonstration program
{\tt demo} in the ``best'' mode, shown in Appendix B.
The printout starts with the detailed specification of the actually used input
parameters. Also, logos of some of the activated subprograms ({\tt KarLud},
{\tt KarFin}) and libraries (\photos, \tauola, \jetset) are printed here.
Next, the printout of one full event (in the standard PDG convention)
is shown (in the actual {\tt demo.output} file, five events can be seen). 
Then, the printouts from the post-generation mode, i.e. 
{\tt CALL YFSWW\_Finalize}, appear.
First, one can see summary reports from the subprograms {\tt KarLud} 
and {\tt KarFin}. Their meaning is rather technical, so the average user
of \yfsww\ does not have to worry about them (unless something unusual
appears there). One thing that the
user may check from time to time there is the entry {\tt B5} in the 
{\tt KarFin} window {\tt B}: the average value of the {\tt WCTRL} weight
should be equal to 1 within the statistical error (if it is not, please notify the
authors!). After these technical printouts from the \yfsww\ subprograms,
there are several summary reports from \tauola\ on the $\tau$ decays. 
The final reports of the main \yfsww\ MC generator are collected in three 
windows: {\tt A}, {\tt B} and {\tt C}. 

The window {\tt A} contains the most important information
from the user's point of view.  
It provides the values of: 
\begin{description}
\item[{\tt  A0}]:  the CMS energy in GeV;
\item[{\tt  A1}]:  the best-order total cross section with its absolute 
                   statistical error for the generated statistics 
                   sample (in pb) ;
\item[{\tt  A2}]:  the relative error of the above cross section;
\item[{\tt  A3}]:  the total number of generated events;
\item[{\tt  A4}]:  the number of accepted events;
\item[{\tt  A5}]:  the number of events with negative weights;
\item[{\tt  A6}]:  the relative contribution to the total cross section 
                   from the events with the negative weights;
\item[{\tt  A7}]:  the number of overweighted events;
\item[{\tt  A8}]:  the relative contribution to the total cross section 
                   from the over-weighted events;
\item[{\tt  A9}]:  the value of the maximum weight for event rejection;
\item[{\tt A10}]:  the average weight.
\end{description}
The numbers in the entries {\tt A5,A6,A7,A8} should be as small as
possible. In particular, the relative contributions to the cross
section from the negative weights and the overweights should be
much smaller than the expected accuracy of the MC calculations.
If there is a large contribution from the overweighted events, one
should try to increase the value of the maximum weight {\tt WtMax},
see Table~\ref{tab:yfsww-input1}. In the case of a large negative weight
contribution, one should contact the authors. The ratio of the
average weight (entry {\tt A10}) to the maximum weight 
(entry {\tt A9}) shows the efficiency of the MC algorithm in the
unweighed events generation, i.e. the event acceptance rate.
In the example given in the {\tt demo} program, it is $\sim 20\%$. 

The window {\tt B} is devoted to the technical information on the ISR
corrections in different orders in $\alpha$ within the YFS framework. 
This information is quite important,  because it shows how big 
the contributions of subsequent orders of the perturbative series are, 
and thus allows us to estimate the missing higher-order effects. 
The entries {\tt B1}--{\tt B4} provide the values of the total cross section 
(in pb) in the orders ${\cal O}(\alpha^0)_{\rm exp}$--${\cal O}(\alpha^3)_{\rm exp}$ 
of the ISR, while the entries  {\tt B15}--{\tt B17}
give the differences of these values in subsequent orders.
The entries {\tt B5}--{\tt B14} contain the information on the contributions
from the individual YFS ISR residuals $\bbeta_{i,ISR}$, also in different
orders in $\alpha$. The differences between various residuals and
various orders for a given residual are provided in the entries 
{\tt B18}--{\tt B26}.

The window {\tt C} contains the information
on the exact ${\cal O}(\alpha^1)_{\rm exp}$ corrections to the $WW$ production
stage. The first entry ({\tt C1}) shows the total ${\cal O}(\alpha^1)_{\rm exp}$
cross section for the exact EW corrections calculation, 
the second one ({\tt C2}) shows the same for the approximate (``fast'')
EW corrections, and the third one ({\tt C3}) contains 
the ${\cal O}(\alpha^1)_{\rm exp}$ LL cross section.
The second part of this window shows the differences between the exact
and approximate EW corrections ({\tt C4)}, and between the exact and the
LL results ({\tt C5)}. The final part (entries {\tt C6}--{\tt C8})
contains the differences of various $\bbeta_i$ contributions.
All these values are given in pb.

This completes the description of the output of \yfsww. 
The remaining entries shown in the {\tt demo.output} file (e.g. histograms)
are produced by the demo main program.

\section{Summary}
We presented the Monte Carlo event generator \yfsww\ version {\tt 1.16}
for the combined $W$-pair production and decay process.
It includes the complete \oal\ EW corrections 
in the $WW$ production process in the leading-pole approximation,
in addition to other numerically sizeable but physically less interesting
effects, such as QED ISR effects up to \Order{\alpha^3}
in the LL approximation.
The program is precise enough to obtain a Standard Model prediction
for the total cross section and any distribution at LEP2, 
understanding that the contribution from the background
diagrams is included in the calculation 
with the help of {\tt KoralW} (or removed from the data).
\yfsww\ includes programming tools to communicate with {\tt KoralW}
in the process of generating a single MC event.
The \yfsww\ program was also tested and is applicable for the energy range up to $1.5\,$TeV
(the \mbox{LC/TESLA} range).
The next possible improvement in the program is the introduction of 
the complete YFS exponentiation
in the MC simulation of the $W$ decays.

\section*{Acknowledgments}

We acknowledge the support of the CERN TH and EP Divisions,
all the LEP Collaborations and the DESY Directorate.
We would like to thank all members of the LEP2 WW/4f Working Group
for many useful discussions,
particularly R.~Chierici, M.~Gr\"unewald, A.~Valassi and
M.~Verzocchi for valuable feedback concerning the development of the code,
and the authors of {\sc RacoonWW} for the useful numerical comparisons.

\newpage
\appendix
\section{Program Parameters and Their Settings}
%
\begin{table}[hp]
\centering
\begin{small}
\begin{tabular}{|l|p{13.0cm}|}
\hline
Parameter & Position and meaning  \\ 
\hline\hline
{\tt CMSEne}        & {\tt xpar(1) (=200.0)}: 
 $\sqrt{s}$, centre-of-mass (CMS) energy  [GeV]\\
{\tt Gmu}           & {\tt xpar(2) (=1.16639d-5)}: 
 $G_F$, Fermi constant [GeV$^{-2}$]\\
{\tt alfWin}        & {\tt xpar(3) (=128.07d0)}:
 $1/\alpha_W$  inverse QED coupling constant at $M_W$ scale\\
{\tt aMaZ}          & {\tt xpar(4) (=91.1882)}:
 $M_Z$, mass  of $Z$ boson, [GeV]\\
{\tt GammZ}         & {\tt xpar(5) (=2.4952)}:
 $\Gamma_Z$, width of $Z$ boson [GeV]\\
{\tt aMaW}          & {\tt xpar(6) (=80.419)}:
 $M_W$, mass  of $W$ boson [GeV]\\
{\tt GammW}         & {\tt xpar(7) (=-2.120)}: 
 $\Gamma_W$, width of $W$ boson [GeV],
 for {(\tt gammw < 0)} $\Gamma_W$ 
 \mbox{\hspace{3.8cm}}
 is recalculated from $G_\mu$, $M_W$ and $\alpha_S$\\
{\tt VVmin}         & {\tt xpar(8) (=1d-6}):
 Minimum $v$-variable (dimensionless), IR cut-off\\
{\tt VVmax}         & {\tt xpar(9) (=0.99)}:
 Maximum value of $v$-variable\\
{\tt WtMax}         & {\tt xpar(10) (=2.0)}:
 Maximum weight for rejection, for {\tt wtmax < 0} redefined
 \mbox{\hspace{3.2cm}} inside the program\\
{\tt aMH}           & {\tt xpar(11) (=115.0)}: Higgs mass [GeV]\\
{\tt aGH}           & {\tt xpar(12) (=1.0)}: Higgs width [GeV]\\
{\tt alpha\_s}      & {\tt xpar(13) (=0.1185)}: QCD coupling constant\\
                    & {\tt xpar(14-19)}: 
 Dummy parameters in \yfsww\ (specific to \koralw) \\
\hline
\end{tabular}
\end{small}
\caption{\sf The list of input parameters of the \yfsww\ generator
             in the {\tt xpar} vector. The default values are in brackets.}
\label{tab:yfsww-input1}
\end{table}
%
\begin{table}[hp]
\label{tab:keycor}
\centering
\begin{small}
\begin{tabular}{|l|p{13.0cm}|}
\hline
Parameter & Position and meaning  \\ 
\hline\hline
{\tt KeyCor}  &  {\tt xpar(2001) (=5)}: Radiative corrections switch\\
              &  {\tt =0}: Born \\
              &  {\tt =1}: Above + ISR\\
              &  {\tt =2}: Above + Coulomb correction\\
              &  {\tt =3}: Above + Full YFS form factor for $WW$ production\\
              &  {\tt =4}: Above + Radiation from $WW$\\
              &  {\tt =5}: Above + Exact \oal\ EW corrections 
                           in $WW$ production (BEST!)\\
              &  {\tt =6}: Same as {\tt 5} but with pretabulated EWC 
                           (approximate but faster!) \\
{\tt KeyLPA}  &  {\tt xpar(2002) (=0)}: LPA mode switch \\
              &  {\tt =0/1}: LPA$_a$/LPA$_b$ 
                 (LPA$_a$ {\bf recommended}, LPA$_b$ for test only) \\
\hline
\end{tabular}
\end{small}
\caption{\sf The list of input parameters in the {\tt xpar} vector 
             specific to the \yfsww\ generator (not used in \koralw). 
             The default values are in brackets.}
\label{tab:yfsww-input2}
\end{table}
%
\begin{table}[hp]
\centering
\begin{small}
\begin{tabular}{|l|p{13.0cm}|}
\hline
Parameter & Position and meaning  \\ 
\hline\hline
                   & {\tt xpar(21-57)}: Values of TGC's: Set 1, 
                     most general set -- 
                     complex numbers, default values are wild random, 
                     not shown\\
{\tt g1(1)   }     & {\tt =DCMPLX(xpar(21),xpar(31))} $=g_1^z$,    
 for $WWZ$ vertex \\
{\tt kap(1)  }     & {\tt =DCMPLX(xpar(22),xpar(32))} $=\kappa_z$, 
 for $WWZ$ vertex \\
{\tt lam(1)  }     & {\tt =DCMPLX(xpar(23),xpar(33))} $=\lambda_z$,
 for $WWZ$ vertex \\
{\tt g4(1)   }     & {\tt =DCMPLX(xpar(24),xpar(34))} $=g_4^z$,    
 for $WWZ$ vertex \\
{\tt g5(1)   }     & {\tt =DCMPLX(xpar(25),xpar(35))} $=g_5^z$,    
 for $WWZ$ vertex \\
{\tt kapt(1) }     & {\tt =DCMPLX(xpar(26),xpar(36))} $=\tilde\kappa_z$, 
 for $WWZ$ vertex \\
{\tt lamt(1) }     & {\tt =DCMPLX(xpar(27),xpar(37))} $=\tilde\lambda_z$,
 for $WWZ$ vertex\\
{\tt g1(2)   }     & {\tt =DCMPLX(xpar(41),xpar(51))} $=g_1^g$,    
 for $WW\gamma$ vertex\\
{\tt kap(2)  }     & {\tt =DCMPLX(xpar(42),xpar(52))} $=\kappa_g$, 
 for $WW\gamma$ vertex\\
{\tt lam(2)  }     & {\tt =DCMPLX(xpar(43),xpar(53))} $=\lambda_g$,
 for $WW\gamma$ vertex\\
{\tt g4(2)   }     & {\tt =DCMPLX(xpar(44),xpar(54))} $=g_4^g$,    
 for $WW\gamma$ vertex\\
{\tt g5(2)   }     & {\tt =DCMPLX(xpar(45),xpar(55))} $=g_5^g$,    
 for $WW\gamma$ vertex\\
{\tt kapt(2) }     & {\tt =DCMPLX(xpar(46),xpar(56))} $=\tilde\kappa_g$,  
 for $WW\gamma$ vertex\\
{\tt lamt(2) }     & {\tt =DCMPLX(xpar(47),xpar(57))} $=\tilde\lambda_g$, 
 for $WW\gamma$ vertex\\
                   & {\tt xpar(61-65)}:  Values of TGCs: Set 2,
                                     see CERN 96-01, Vol. 1, p. 525 \\
{\tt  delta\_Z }   & {\tt =xpar(61)} $=\delta_Z$ \\ 
{\tt  x\_gamma }   & {\tt =xpar(62)} $=x_{\gamma}$ \\
{\tt  x\_Z     }   & {\tt =xpar(63)} $=x_Z$ \\
{\tt  y\_gamma }   & {\tt =xpar(64)} $=y_{\gamma}$ \\
{\tt  y\_Z     }   & {\tt =xpar(65)} $=y_Z$\\
                   & {\tt xpar(71-73)}: Values of TGCs: Set 3,
                                     see CERN 96-01, Vol. 1, p. 525\\
{\tt alpha\_Wphi}  & {\tt =xpar(71)} $=\alpha_{W\phi}$\\
{\tt alpha\_Bphi}  & {\tt =xpar(72)} $=\alpha_{B\phi}$\\
{\tt alpha\_W   }  & {\tt =xpar(73)} $=\alpha_W$\\
{\tt amel  }       & {\tt =xpar(100) (=0.510998902d-3)}: beam (electron) mass\\
{\tt AlfInvl  }    & {\tt =xpar(101) (=137.03599976)}:
 $1/\alpha_{\rm QED}$ at the Thomson limit \\
{\tt gpicob  }     & {\tt =xpar(101) (=389.379292d6)}:
 GeV$^{-2}$ to picobarn translation constant\\
{\tt BR(1:20)}      &  {\tt xpar(131-139)}:
 $W$ branching ratios (BR); the numbering of entries is:\\
&
 $1=ud,\; 2=cd,\; 3=us,\; 4=cs,\; 5=ub,\; 6=cb,\; 7=e\nu_e,\; 8=\mu\nu_{\mu},
 \; 9=\tau\nu_{\tau}$\\
{\tt amafin(20)}    &  {\tt xpar(500 +10*KF +6)}:
 Masses of the $W$ decay products; the used entries KF 
 \mbox{\hspace{4.1cm}}
 (in PDG notation) are:\\
& \mbox{\hspace{4.1cm}}
 $1=d,\;  2=u,\;  3=s,\;  4=c,\;  5=b,\; 6=t$,\\
& \mbox{\hspace{4.1cm}}
 $11=e,\; 12=\nu_e,\; 13=\mu,\; 14=\nu_\mu,\; 15=\tau,\; 16=\nu_\tau$.\\
&
 Note: The masses of $\tau$ and $\nu_\tau$ have to be independently 
       set to the same \mbox{\hspace{2mm}} \mbox{\hspace{11mm}}
       numerical values in the initialization of \tauola\\
{\tt VCKM(1:3,1:3)} &  {\tt xpar(111-119)}:
                       CKM matrix elements (PDG 2000)\\
\hline
\end{tabular}
\end{small}
\caption{\sf The list of input parameters of the \yfsww\ generator
             in the {\tt xpar} vector (cont.). The default values are in brackets.}
\label{tab:yfsww-input3}
\end{table}
%
\begin{table}[hp]
\centering
\begin{small}
\begin{tabular}{|l|p{13.0cm}|}
\hline
Parameter & Position and meaning  \\ 
\hline\hline
              &  {\tt xpar(1011-1013)}:
 Dummy paremeters in \yfsww\ (specific to \koralw)\\
{\tt KeyCul}  &  {\tt xpar(1014) (=2)}: Coulomb correction switch\\
              &  {\tt =0}: Coulomb correction is OFF\\
              &  {\tt =1}: ``Standard'' Coulomb correction is ON\\
              &  {\tt =2}: ``Screened'' Coulomb correction is ON\\
{\tt KeyBra}  &  {\tt xpar(1021) (=2)}: Sets $W$ branching ratios (BR),
 used for normalization of \mbox{\hspace{3.2cm}} the matrix element\\
              &  {\tt =0}: Born values (no mixing) with ``naive'' QCD
                 (if $\alpha_S=0$: 
 $ud, cs = 1/3$,\; \mbox{\hspace{0.5cm}} \mbox{\hspace{0.5cm}}
 $e\nu,\mu\nu,\tau\nu=1/9$,\; others $=0$)\\
              &  {\tt =1}: Values of BRs taken from the input\\
              &  {\tt =2}: With CKM mixing and ``naive'' QCD,
                          calculated in the IBA from the \mbox{\hspace{2mm}} 
                          \mbox{\hspace{0.6cm}} CKM matrix (PDG 2000)\\
{\tt KeyMas}  &  {\tt xpar(1022)}:
 Dummy parameter in \yfsww\ (specific to \koralw)\\
{\tt KeyZet}  &  {\tt xpar(1023) (=1)}: $Z$ width choice\\
              &  {\tt =0}: $Z$ width in $Z$ propagator: $(s/M_Z) \Gamma_Z$\\
              &  {\tt =1}: $Z$ width in $Z$ propagator:   $M_Z \Gamma_Z$\\
              &  {\tt =2}: no $Z$ width in $Z$ propagator\\
{\tt KeySpn}  &  {\tt xpar(1024) (=1)}: Spin effects in $W$ decays\\
              &  {\tt =0}: Spin effects OFF (for tests only)\\
              &  {\tt =1}: Spin effects ON (recommended)\\
{\tt KeyRed}  &  {\tt xpar(1025)}:
                 Dummy paremeter in \yfsww\ (specific to \koralw)\\
{\tt KeyWu}   &  {\tt xpar(1026) (=1)}: $W$ width choice\\
              &  {\tt =0}: $W$ width in $W$ propagator: $(s/M_W) \Gamma_W$\\
              &  {\tt =1}: $W$ width in $W$ propagator: $M_W \Gamma_W$\\
              &  {\tt =2}: No $W$ width ($\Gamma_W=0$) in $W$ propagator 
                           (dangerous!)\\
{\tt KeyWgt}  &  {\tt xpar(1031) (=0)}: Unweighted/Weighted choice\\
              &  {\tt =0}: Unweighted events {\tt WtMod}=1,
 for apparatus Monte Carlo\\
              &  {\tt =1}: Weighted events (variable weights)\\
{\tt KeyRnd}  &  {\tt xpar(1032) (=1)}:  Random number generator switch\\
              &  {\tt =1}: RANMAR random number generator\\ 
              &  {\tt =2}: ECURAN random number generator\\ 
              &  {\tt =3}: CARRAN random number generator\\ 
{\tt KeySmp}  &  {\tt xpar(1033)}:
 Dummy paremeter in \yfsww\ (specific to \koralw)\\
{\tt KeyMix}  &  {\tt xpar(1041) (=1)}: Input parameter scheme (IPS) switch\\
              &  {\tt =0}: LEP2 '95 Workshop scheme 
                 (only for Born and ISR)\\
              &  {\tt =1}: $G_\mu$-scheme
                  ({\bf recommended}; 
                   the only choice for \oal\ EW corrections)  \\
              &  {\tt =2}: $\alpha$-scheme (``pure'' Born; for tests)\\
{\tt Key4f}   &  {\tt xpar(1042)}:
                 Dummy parameter in \yfsww\ (specific to \koralw)\\
\hline
\end{tabular}
\end{small}
\caption{\sf The list of input parameters of the \yfsww\ generator
             in the {\tt xpar} vector (cont.). The default values are in brackets.}
\label{tab:yfsww-input4}
\end{table}
%
\begin{table}[hp]
\centering
\begin{small}
\begin{tabular}{|l|p{13.0cm}|}
\hline
Parameter & Position and meaning  \\ 
\hline\hline
{\tt KeyAcc}  &  {\tt xpar(1043) (=0)}: Anomalous TGCs switch\\
              &  {\tt =0}: Anomalous $WWV$ couplings OFF\\
              &  {\tt >0}: Anomalous $WWV$ couplings ON\\
              &  {\tt =1}:
 The most general (complex number) TGCs in the notation
 of Hagiwara \mbox{\hspace{0.6cm}}
 et al., Nucl. Phys. {\bf B282} (1987) 253\\
              &  {\tt =2} Parametrization of CERN 96-01,
                        Vol. 1, p. 525:
 $\delta_Z, x_\gamma, x_Z, y_\gamma, y_Z$\\
              &  {\tt =3}: Parametrization of CERN 96-01,
 Vol. 1, p. 525: $\alpha_{W\phi}, \alpha_{B\phi}, \alpha_W$\\
{\tt KeyZon}  &  {\tt xpar(1044)}:
 Dummy parameter in \yfsww\ (specific to \koralw)\\
{\tt KeyWon}  &  {\tt xpar(1045)}:
 Dummy parameter in \yfsww\ (specific to \koralw)\\
{\tt KeyDwm}    &  {\tt xpar(1055) (=0)}:  Sets decay channel of $W^-$ \\
                &  {\tt =0}: Inclusive; otherwise the exclusive modes for the $W$ are:\\
                &  
 $\begin{bmatrix}  =1  &  2 &  3 &  4 &  5 &  6 &    7 &      8 &     9\\ 
                    ud & cd & us & cs & ub & cb & e\nu & \mu\nu & \tau\nu 
                    
 \end{bmatrix}$ \\
{\tt KeyDwp}    &  {\tt xpar(1056) (=0)}: Sets the decay channel of $W^+$,
                   similar assignments as for \mbox{\hspace{3.2cm}} $W^-$\\
{\tt Nout}      &  {\tt xpar(1057) (=-1)}:
  Output unit number for the generator
 (if $<0$ then \mbox{\hspace{3.6cm}} {\tt Nout=16})\\
{\tt Jak1}      &  {\tt xpar(1071) (=0)}:
 Input for \tauola, defines decay mode of the $\tau^+$ 
 in $W^+$ \mbox{\hspace{3.2cm}} decay\\
{\tt Jak2}      &  {\tt xpar(1072) (=0)}:
 Input for \tauola, defines decay mode of the $\tau^-$
 in $W^-$ \mbox{\hspace{3.2cm}} decay\\
                &  {\tt Jak1,Jak2 = -1}: \tauola\ is switched OFF\\
                &  {\tt Jak1,Jak2 = 0}: Requests all $\tau^\pm$ decay
 channels to be simulated\\ 
                &  {\tt Jak1,Jak2 > 0}: Single specific $\tau^\pm$ decay
 channel, see \tauola\ manual\\
{\tt Itdkrc}    &  {\tt xpar(1073) (=1)}:
 Input  for \tauola, radiative corrections in leptonic
 $\tau$ decays \mbox{\hspace{3.2cm}} switch\\
                &  {\tt Itdkrc = 1}: Corrections are  ON\\
                &  {\tt Itdkrc = 0}: Corrections are  OFF\\
{\tt IfPhot}    &  {\tt xpar(1074) (=1)}: \photos\ activation switch\\
                &   {\tt =0}: \photos\ is OFF\\
                &   {\tt =1}: Radiation in leptonic $W$ decays is ON\\
                &   {\tt =2}: Radiation in both leptonic and quarkonic
 $W$ decays is ON (for tests)\\
{\tt IfHadM}    &  {\tt xpar(1075) (=1)}: $W^-$ hadronization activation
 switch (\jetset)\\
{\tt IfHadP}    &  {\tt xpar(1075) (=1)}: $W^+$ hadronization activation
 switch (\jetset)\\
                &  {\tt IfHadM,ifhadP=0}: Hadronization is OFF\\
                &  {\tt IfHadM,ifhadP=1}: Hadronization is ON\\
                & \mbox{\hspace{3.2cm}}
 In the present version {\tt IfHadM} and {\tt IfHadP} have to be equal!\\
{\tt Umask}     &  {\tt xpar(1101-1302)}:
 Dummy parameters in \yfsww\ (specific to \koralw)\\
 BE params      &  {\tt xpar(4061-4070)}:
 Dummy parameters in \yfsww\ (specific to \koralw)\\
\hline
\end{tabular}
\end{small}
\caption{\sf The list of input parameters of the \yfsww\ generator
             in the {\tt xpar} vector (cont.). The default values are in brackets.}
\label{tab:yfsww-input5}
\end{table}
%
\begin{table}[hp]
\centering
\begin{small}
\begin{tabular}{|l|p{13.0cm}|}
\hline
Parameter & Meaning  \\ \hline\hline
{\tt Q1(4)}     & Four-momentum of $W^-$\\
{\tt Q2(4)}     & Four-momentum of $W^+$\\
{\tt SPhum(4)}  & Sum of four-momenta of photons from the $WW$ production stage\\
{\tt SPhot(100,4)}  & Four-momenta of photons from the $WW$ production stage \\
{\tt Nphot}        & Number of photons from the $WW$ production stage\\
\hline
\end{tabular}
\end{small}
\caption{\sf The list of four-momenta in the {\tt COMMON /MOMWWP/}
             of the \yfsww\ generator. 
             They are given in {\rm GeV} in the CMS of the incoming beams.
        }
\label{tab:yfsww-momwwp}
\end{table}
%
\begin{table}[hp]
\centering
\begin{small}
\begin{tabular}{|l|p{13.0cm}|}
\hline
Parameter & Meaning  \\
\hline\hline
{\tt pf1(4)} & Four-momentum of the fermion from the $W^-$ decay\\
{\tt pf2(4)} & Four-momentum of the antifermion from the $W^-$ decay\\
{\tt pf3(4)} & Four-momentum of the fermion from the $W^+$ decay\\
{\tt pf4(4)} & Four-momentum of the antifermion from the $W^+$ decay\\
{\tt Phot(100,4)}  & Four-momenta of photons from the $WW$ production stage\\
{\tt Npho}        & Multiplicity of photons from the $WW$ production stage\\
\hline
\end{tabular}
\end{small}
\caption{\sf The list of four-momenta in the {\tt COMMON /MOMDEC/}
             of the \yfsww\ generator.
             They are given in {\rm GeV} in the CMS of the incoming beams.
        }
\label{tab:yfsww-momdec}
\end{table}
%
\begin{table}[hp]
\centering
\begin{small}
\begin{tabular}{|l|l|p{8.5cm}|}
\hline
Routine & Parameter & Meaning  \\ 
\hline\hline
YFSWW\_GetBeams & {\tt q1(4)} & Four-momentum of the $e^-$ beam\\
                & {\tt q2(4)} & Four-momentum of the $e^+$ beam\\
\hline
YFSWW\_Get4f &
{\tt flav(4)} & Flavours of the final $4f$-state in the PDG convention\\
& 
{\tt p1(4)} & Four-momentum of the fermion of {\tt flav(1)}\\
&
{\tt p2(4)} & Four-momentum of the fermion of {\tt flav(2)}\\
&
{\tt p3(4)} & Four-momentum of the fermion of {\tt flav(3)}\\
&
{\tt p4(4)} & Four-momentum of the fermion of {\tt flav(4)}\\
\hline
YFSWW\_GetPhotAll &
{\tt NphAll}        & Number of photons from the WW production stage\\
&
{\tt PhoAll(100,4)} & Four-momenta of these photons\\
\hline
\end{tabular}
\end{small}
\caption{\sf The list of output parameters of the \yfsww\ generator transferred
         through parameters of the getter-routines: 
         {\tt YFSWW\_GetBeams(q1,q2)},  
         {\tt YFSWW\_Get4f(flav,p1,p2,p3,p4)} and 
         {\tt YFSWW\_GetPhotAll(NphAll,PhoAll)}.
        }
\label{tab:yfsww-event}
\end{table}
%
\begin{table}[hp]
\centering
\begin{small}
\begin{tabular}{|l|p{11.0cm}|}
\hline
Parameter & Meaning  \\ 
\hline\hline
{\tt WtCru1}  & Crude weight for ISR \\
{\tt WtCru2}  & Crude weight for the rest\\ 
              & They are necessary to build the total weight out of {\tt WtSet}\\
{\tt WtMod}   & The ``best'' weight corresponding to Eq.~(\ref{Master}), e.g. for {\tt KeyCor=5}:\\ 
              & {\tt WtMod = Wtcru1*Wtcru2*(WtSet(41)-WtSet(2)+WtSet(4))}\\
\hline
{\tt WtSet(1-19)} & Various ISR-type contributions to the matrix element 
 ($\bbeta$ functions) -- to get the total weight they must be multiplied by 
 {\tt WtCru1} \\
{\tt WtSet(1)}     & Zeroth-order ($\bbeta_0$)\\
{\tt WtSet(2)}     & First-order ISR ($\bbeta_0 +\bbeta_1$)\\
{\tt WtSet(3)}     & Second-order ISR ($\bbeta_0  +\bbeta_1 +\bbeta_2$)\\
{\tt WtSet(4)}     & Third-order ISR
 ($\bbeta_0  +\bbeta_1 +\bbeta_2 +\bbeta_3$)\\
{\tt WtSet(10)}    & ${\cal O}(\alpha^0)$ contribution from $\bbeta_0$\\
{\tt WtSet(11)}    & ${\cal O}(\alpha^1)$ ISR contribution from $\bbeta_0$ \\
{\tt WtSet(12)}    & ${\cal O}(\alpha^1)$ ISR contribution from $\bbeta_1$ \\
{\tt WtSet(13)}    & ${\cal O}(\alpha^2)$ ISR contribution from $\bbeta_0$ \\ 
{\tt WtSet(14)}    & ${\cal O}(\alpha^2)$ ISR contribution from $\bbeta_1$ \\ 
{\tt WtSet(15)}    & ${\cal O}(\alpha^2)$ ISR contribution from $\bbeta_2$ \\ 
{\tt WtSet(16)}    & ${\cal O}(\alpha^3)$ ISR contribution from $\bbeta_0$ \\
{\tt WtSet(17)}    & ${\cal O}(\alpha^3)$ ISR contribution from $\bbeta_1$ \\ 
{\tt WtSet(18)}    & ${\cal O}(\alpha^3)$ ISR contribution from $\bbeta_2$ \\  
{\tt WtSet(19)}    & ${\cal O}(\alpha^3)$ ISR contribution from $\bbeta_3$ \\  
\hline
{\tt WtSet(41-52)} & Contributions to the matrix element from \oal\ 
 EW corrections in the $WW$ production stage -- 
 to get the total weight they must be multiplied by {\tt WtCru1*WtCru2}\\
{\tt WtSet(41)}    & ${\cal O}(\alpha^1)$ EW corrections
 ($\bbeta_0 +\bbeta_1$)\\
{\tt WtSet(51)}    & Pretabulated (``fast'')  ${\cal O}(\alpha^1)$ 
                     EW corrections ($\bbeta_0 +\bbeta_1$)\\
{\tt WtSet(42)}    & ${\cal O}(\alpha^1)$ contribution from $\bbeta_0$ \\
{\tt WtSet(43)}    & ${\cal O}(\alpha^1)$ contribution from $\bbeta_1$ \\
{\tt WtSet(52)}    & Pretabulated ${\cal O}(\alpha^1)$ contribution 
                     from $\bbeta_0$ \\
\hline
{\tt WtSet(65-68)} & Fixed-order (no exponentiation) contributions 
                     (for tests)\\
{\tt WtSet(65)}    & ${\cal O}(\alpha^0)$, no exponentiation\\
{\tt WtSet(66)}    & ${\cal O}(\alpha^1)$, no exponentiation\\
{\tt WtSet(67)}    & ${\cal O}(\alpha^1)$ virtual + soft real photon only, 
                     no exponentiation\\
{\tt WtSet(68)}    & ${\cal O}(\alpha^1)$ real hard photon only, 
                     no exponentiation\\
\hline
\end{tabular}
\end{small}
\caption{\sf The list of output weights in the {\tt COMMON /WGTALL/}
             of the \yfsww\ generator.}
\label{tab:yfsww-weights}
\end{table}
%
\begin{table}[hp]
\centering
\begin{small}
\begin{tabular}{|l|l|p{9.0cm}|}
\hline
Routine & Parameter & Meaning  \\ 
\hline\hline
{\tt YFSWW\_GetWtMain} &
{\tt WtMain}  & {\it Principal} best MC event weight\\
\hline
{\tt YFSWW\_GetWtISR} &
{\tt WtISR}& Best ISR-type weight \\
\hline
{\tt YFSWW\_GetWtAll} &
{\tt WtMain}   & {\it Principal} best MC event weight\\
&
{\tt WtCrud}   & Crude MC event weight: {\tt WtCrud=WtCru1*WtCru2} \\
&
{\tt WtSetAll(100)} & Model weights \\
&  & (the same as in {\tt WtSet(100)} in {\tt COMMON /WGTALL/})\\
\hline
\end{tabular}
\end{small}
\caption{\sf The list of output parameters of the \yfsww\ generator transferred 
             through parameters of the getter-routines: 
             {\tt YFSWW\_GetWtMain(WtMain)}, 
             {\tt YFSWW\_GetWtISR(WtISR)} and 
             {\tt YFSWW\_GetWtAll(WtMain,WtCrud,WtSetAll)}.
        }
\label{tab:yfsww-getwgt}
\end{table}
\begin{table}[hp]
\centering
\begin{small}
\begin{tabular}{|l|l|p{9.0cm}|}
\hline
Routine & Parameter & Meaning  \\ 
\hline\hline
{\tt YFSWW\_GetXSecMC} &
{\tt XSecMC}  & {\it Principal} total MC cross section [pb]\\
&
{\tt XErrMC}  & Its absolute error [pb]\\
\hline
{\tt YFSWW\_GetXSecNR} &
{\tt XSecNR}& Normalization (primary) cross section, see Sect.~\ref{sec:inp-outp}: \\
&
         & For {\tt KeyWgt=0}, the principal cross section {\tt XSecMC} [pb],\\
&
         & For {\tt KeyWgt=1}, the crude cross section {\tt XCrude} [pb]\\
&
{\tt XErrNR} & Its absolute error [pb]\\
\hline
{\tt YFSWW\_GetNevMC} &
{\tt NevMC}   & Total number of generated Monte Carlo events\\
\hline
\end{tabular}
\end{small}
\caption{\sf The list of output parameters of the \yfsww\ generator transferred 
             through parameters of the getter-routines: 
             {\tt YFSWW\_GetXSecMC(XSecMC,XErrMC)}, 
             {\tt YFSWW\_GetXSecNR(XSecNR,XErrNR)} and 
             {\tt YFSWW\_GetNevMC(NevMC)}.
        }
\label{tab:yfsww-final}
\end{table}
\vfill

\newpage
\section{Output of the Demo Program}
\label{printout}

{
\small
\renewcommand{\baselinestretch}{0.4}
\begin{verbatim}
  

 ***************************************************************************
 *             YFSWW3 version 1.16,  January 2001                          *
 *             S. Jadach, W. Placzek, M. Skrzypek,                         *
 *                     B.F.L. Ward, Z. Was                                 *
 *                                                                         *
 *                        INPUT PARAMATERS:                                *
 *     200.00000000                 CMS energy total         CMSEne    I.0 *
 * ***********************************************                         *
 *                5                 Rad. Corr. switch        KeyCor    IR1 *
 * ***********************************************                         *
 *             2101                 QED super-switch         KeyRad    IQ1 *
 *                1                 Init. state Rad.         KeyISR    IQ2 *
 *                0                 Final state Rad.         KeyFSR    IQ3 *
 *                1                 Next-To-Leading          KeyNLL    IQ4 *
 *                2                 Coulomb corr.            KeyCul    IQ5 *
 * ***********************************************                         *
 *           101112                 Physics super-switc      KeyPhy    IP1 *
 *                0                 FS mass reduction        KeyRed    IP2 *
 *                1                 Spin in W decays         KeySpn    IP3 *
 *                1                 Z propag.                KeyZet    IP4 *
 *                1                 Mass kinematics.         KeyMas    IP5 *
 *                2                 Branching Rat.           KeyBra    IP6 *
 *                1                 W propag.                KeyWu     IP7 *
 * ***********************************************                         *
 *               10                 Technical super-swi      KeyTek    IT1 *
 *                0                 LPA_a or LPA_b mode      KeyLPA    IT2 *
 *                0                 Presampler type          KeySmp    IT3 *
 *                1                 Rand. Numb. Gen.         KeyRnd    IT4 *
 *                0                 Weighting  switch        KeyWgt    IT5 *
 * ***********************************************                         *
 *            10001                 Miscelaneous             KeyMis    IM1 *
 *                1                 EW Input Par. Schem      KeyMix    IM2 *
 *                0                 4 fermion matr el        Key4f     IM3 *
 *                0                 Anomalous couplings      KeyAcc    IM4 *
 *                1                 WW type final state      KeyWon    IM5 *
 *                0                 ZZ type final state      KeyZon    IM6 *
 * ***********************************************                         *
 *                0                 W- decay mode            KeyDWm    ID1 *
 *                0                 W+ decay mode            KeyDWp    ID2 *
 * ***********************************************                         *
 *       1.16639000                 G_mu * 1d5               Gmu       I.1 *
 *     132.50494581                 Inverse alpha_w          alfWin    I.2 *
 *      91.18820000                 Z mass   [GeV]           aMaZ      I.3 *
 *       2.49520000                 Z width  [GeV]           GammZ     I.4 *
 *      80.41900000                 W mass   [GeV]           aMaW      I.5 *
 *       2.09957845                 W width  [GeV]           GammW     I.6 *
 *       0.00000100                 Dummy infrared cut       VVmin     I.7 *
 *       0.99000000                 v_max ( =1 )             VVmax     I.8 *
 *       4.00000000                 Max wt for rejectn.      WtMax     I.9 *
 *       4.00000000                 Max wt for CC03 rej      WtMax     I10 *
 *       0.11850000                 alpha_s: QCD coupl.      alphas    I11 *
 *       0.00000000                 Color Re-Con. Prob.      PReco     I12 *
 * ***********************************************                         *
 *       0.22224994                 sin(theta_W)**2          sinW2     I13 *
 * ***********************************************                         *
 * ***********************************************                         *
 *         Z width in Z propagator:   M_Z *GAMM_Z                          *
 * ***********************************************                         *
 *                                                                         *
 *                            CKM matrix elements:                         *
 *       0.97493000                                V_ud   VCKM(1,1)    IV1 *
 *       0.22250000                                V_us   VCKM(1,2)    IV2 *
 *       0.00350000                                V_ub   VCKM(1,3)    IV3 *
 *      -0.22246000                                V_cd   VCKM(2,1)    IV4 *
 *       0.97412000                                V_cs   VCKM(2,2)    IV5 *
 *       0.04000000                                V_cb   VCKM(2,3)    IV6 *
 *       0.00549000                                V_td   VCKM(3,1)    IV7 *
 *      -0.03978000                                V_ts   VCKM(3,2)    IV8 *
 *       0.99920000                                V_tb   VCKM(3,3)    IV9 *
 *              Unitarity check of the CKM matrix:                         *
 *                       1.00001   0.00000   0.00000                       *
 *               VV+ =   0.00000   1.00000   0.00000                       *
 *                       0.00000   0.00000   1.00001                       *
 *                                                                         *
 *                   Branching ratios of W decays:                         *
 *       0.32071536                                  ud       BR(1)    IB1 *
 *       0.01669847                                  cd       BR(2)    IB2 *
 *       0.01670448                                  us       BR(3)    IB3 *
 *       0.32018266                                  cs       BR(4)    IB4 *
 *       0.00000413                                  ub       BR(5)    IB5 *
 *       0.00053987                                  cb       BR(6)    IB6 *
 *       0.10838559                                   e       BR(7)    IB7 *
 *       0.10838559                                  mu       BR(8)    IB8 *
 *       0.10838559                                 tau       BR(9)    IB9 *
 *                                 fermion masses:                         *
 *       0.00600000                                   d   amafin(1)    IM1 *
 *       0.00300000                                   u   amafin(2)    IM2 *
 *       0.12250000                                   s   amafin(3)    IM3 *
 *       1.25000000                                   c   amafin(4)    IM4 *
 *       4.20000000                                   b   amafin(5)    IM5 *
 *     174.30000000                                   t   amafin(6)    IM6 *
 *       0.00051100                                   e  amafin(11)    IM7 *
 *       0.00000000                                  ve  amafin(12)    IM8 *
 *       0.10565836                                  mu  amafin(13)    IM9 *
 *       0.00000000                                 vmu  amafin(14)   IM10 *
 *       1.77703000                                 tau  amafin(15)   IM11 *
 *       0.00000000                                vtau  amafin(16)   IM12 *
 *                          Higgs mass:                                    *
 *     115.00000000                               Higgs       aMHig    IMH *
 *                                                                         *
 *                                 DECAY LIBRARIES                         *
 *                0                       TAUOLA for W+        Jak1    IL1 *
 *                0                       TAUOLA for W-        Jak2    IL2 *
 *                1                   TAUOLA Ord(alpha)      Itdkrc    IL3 *
 *                1                              PHOTOS      IfPhot    IL4 *
 *                1                       JETSET for W-      IfHadM    IL5 *
 *                1                       JETSET for W+      IfHadP    IL6 *
 ***************************************************************************
 


            *************************************************************
            *  ###   ### ########  ######  ###   ### ###   ###   #####  *
            *  ###   ### ######## ######## ###   ### ###   ###  ####### *
            *   ### ###  ###      ####     ###   ### ###   ###      ### *
            *     ###    ######     ####   ###   ### ###   ###    ###   *
            *     ###    ######       #### ### # ### ### # ###      ### *
            *     ###    ###      ######## #### #### #### ####  ####### *
            *     ###    ###       ######  ###   ### ###   ###   #####  *
            *************************************************************


           
 ***************************************************************************
 *                        Initialize KarLud  start                         *
 *      19.33542764                 xs_crude  VESKO          XCVESK        *
 *      19.19161478                 xs_crude  GAUSS          XCGAUS        *
 *       0.00749353                 XCVESK/XCGAUS-1                        *
 *                        Initialize KarLud  end                           *
 ***************************************************************************



 ***************************************************************************
 *                        Initialize KarFin  start                         *
 *                        Initialize KarFin  end                           *
 ***************************************************************************



 ***************************************************************************
 *                         *****TAUOLA LIBRARY: VERSION 2.6 ******         *
 *                         ***********August   1995***************         *
 *                         **AUTHORS: S.JADACH, Z.WAS*************         *
 *                         **R. DECKER, M. JEZABEK, J.H.KUEHN*****         *
 *                         **AVAILABLE FROM: WASM AT CERNVM ******         *
 *                         ***** PUBLISHED IN COMP. PHYS. COMM.***         *
 *                         *******CERN-TH-5856 SEPTEMBER 1990*****         *
 *                         *******CERN-TH-6195 SEPTEMBER 1991*****         *
 *                         *******CERN-TH-6793 NOVEMBER  1992*****         *
 *                         **5 or more pi dec.: precision limited          *
 *                         ******DEXAY ROUTINE: INITIALIZATION****         *
 *                   0     JAK1   = DECAY MODE FERMION1 (TAU+)             *
 *                   0     JAK2   = DECAY MODE FERMION2 (TAU-)             *
 ***************************************************************************



  --------------- YFSWW demo -------------------
  NevTot =  1000 events to be generated ...
\end{verbatim}
}
{
\tiny
\begin{verbatim}
   ..... skipped ....

                            Event listing (standard)

    I  particle/jet  K(I,1)   K(I,2) K(I,3)     K(I,4)      K(I,5)       P(I,1)       P(I,2)       P(I,3)       P(I,4)       P(I,5)

    1  !e-!              21       11     0           3           4      0.00000      0.00000    100.00000    100.00000      0.00051
    2  !e+!              21      -11     0           3           4      0.00000      0.00000   -100.00000    100.00000      0.00051
    3  (W-)              11      -24     1           5           6    -58.15061      0.61540     33.01304     95.19679     67.75468
    4  (W+)              11       24     1           7           8     58.15061     -0.61540    -33.01304    104.80321     80.69684
    5  (d)               14        1     3   3   6  10   0   0  10    -62.20078    -14.63227      6.34668     64.21308      0.00600
    6  (u~)              14       -2     3   0   0  11   3   5  11      4.05017     15.24767     26.66636     30.98371      0.00300
    7  (u)               14        2     4   3   8  29   0   0  29     18.03726      6.52502     26.47908     32.69649      0.00300
    8  (d~)              14       -1     4   0   0  30   3   7  30     40.11335     -7.14042    -59.49212     72.10672      0.00600
    9  (CMshower)        11       94     5          10          11    -58.15062      0.61540     33.01304     95.19679     67.75468
   10  (d)               14        1     9   3   5  13   0   5  12    -60.91610    -14.19066      6.47966     63.22975      6.62341
   11  (u~)              14       -2     9   0   6  14   3   6  15      2.76548     14.80606     26.53338     31.96705      9.53966
   12  (d)               14        1    10   3  13  17   0  10  16    -46.33492     -8.95829      6.22708     47.67010      2.54677
   13  (g)               14       21    10   3  10  18   3  12  19    -14.58117     -5.23237      0.25259     15.55965      1.43197
   14  (u~)              14       -2    11   0  11  20   3  15  21      2.79517     14.95296     25.98681     31.40030      8.90286
   15  (g)               13       21    11   2  14   0   2  11   0     -0.02970     -0.14691      0.54657      0.56675      0.00000
   16  (d)               13        1    12   2  17   0   0  12   0    -12.64723     -1.30492      1.82179     12.84423      0.00990
   17  (g)               13       21    12   2  12   0   2  16   0    -33.68769     -7.65337      4.40528     34.82587      0.00000
   18  (g)               13       21    13   2  13   0   2  19   0     -9.68361     -4.03725     -0.24096     10.49428      0.00000
   19  (g)               13       21    13   2  18   0   2  13   0     -4.89756     -1.19512      0.49355      5.06537      0.00000
   20  (u~)              14       -2    14   0  14  22   3  21  23      4.17251     13.40366     18.62694     23.57434      3.42331
   21  (g)               14       21    14   3  20  24   3  14  25     -1.37734      1.54930      7.35988      7.82596      1.66749
   22  (u~)              14       -2    20   0  20  26   3  23  27      4.29870     12.59084     16.24373     21.05317      1.53932
   23  (g)               13       21    20   2  22   0   2  20   0     -0.12619      0.81281      2.38321      2.52117      0.00000
   24  (g)               13       21    21   2  21   0   2  25   0     -1.25460      0.15914      3.44828      3.67287      0.00000
   25  (g)               13       21    21   2  24   0   2  21   0     -0.12274      1.39016      3.91159      4.15309      0.00000
   26  (u~)              13       -2    22   0  22   0   2  27   0      2.89930      7.50857      8.72410     11.86990      0.00560
   27  (g)               13       21    22   2  26   0   2  22   0      1.39939      5.08227      7.51962      9.18327      0.00000
   28  (CMshower)        11       94     7          29          30     58.15061     -0.61540    -33.01304    104.80321     80.69683
   29  (u)               14        2    28   3   7  32   0   7  31     23.86922      5.09816     15.70427     43.16896     31.95629
   30  (d~)              14       -1    28   0   8  33   3   8  34     34.28139     -5.71356    -48.71731     61.63424     14.74939
   31  (u)               13        2    29   2  32   0   0  29   0      2.06978     -3.05418     -5.61577      6.71930      0.00560
   32  (g)               14       21    29   3  29  35   3  31  36     21.79944      8.15234     21.32004     36.44967     18.23068
   33  (d~)              14       -1    30   0  30  37   3  34  38      9.91990      3.71419    -14.20958     17.89177      2.45019
   34  (g)               14       21    30   3  33  39   3  30  40     24.36149     -9.42775    -34.50773     43.74247      6.34476
   35  (g)               14       21    32   3  32  41   3  36  42     24.19235      8.81956     20.09323     33.02196      4.86387
   36  (g)               14       21    32   3  35  43   3  32  44     -2.39291     -0.66722      1.22681      3.42771      2.01816
   37  (d~)              13       -1    33   0  33   0   2  38   0      7.66346      2.82714    -12.47235     14.90909      0.00990
   38  (g)               13       21    33   2  37   0   2  33   0      2.25644      0.88705     -1.73723      2.98268      0.00000
   39  (g)               14       21    34   3  34  45   3  40  46     11.13057     -1.65836    -12.38914     16.78682      1.29097
   40  (g)               14       21    34   3  39  47   3  34  48     13.23092     -7.76938    -22.11859     26.95566      1.39821
   41  (g)               13       21    35   2  35   0   2  42   0      6.51083      2.61590      3.06918      7.65857      0.00000
   42  (g)               14       21    35   3  41  49   3  35  50     17.68152      6.20366     17.02405     25.36338      1.53671
   43  (g)               13       21    36   2  36   0   2  44   0     -2.15456     -0.80571      0.13785      2.30441      0.00000
   44  (g)               13       21    36   2  43   0   2  36   0     -0.23835      0.13849      1.08896      1.12330      0.00000
   45  (g)               13       21    39   2  39   0   2  46   0      9.54865     -1.43393     -9.90045     13.82938      0.00000
   46  (g)               13       21    39   2  45   0   2  39   0      1.58193     -0.22443     -2.48869      2.95744      0.00000
   47  (g)               13       21    40   2  40   0   2  48   0      1.41318     -1.20728     -3.26543      3.75734      0.00000
   48  (g)               13       21    40   2  47   0   2  40   0     11.81774     -6.56210    -18.85316     23.19831      0.00000
   49  (g)               13       21    42   2  42   0   2  50   0     14.75801      5.72986     14.60889     21.54181      0.00000
   50  (g)               13       21    42   2  49   0   2  42   0      2.92351      0.47380      2.41516      3.82157      0.00000
   51  (d)           A   12        1    16          73          73    -12.64723     -1.30492      1.82179     12.84423      0.00990
   52  (g)           I   12       21    17          73          73    -33.68769     -7.65337      4.40528     34.82587      0.00000
   53  (g)           I   12       21    19          73          73     -4.89756     -1.19512      0.49355      5.06537      0.00000
   54  (g)           I   12       21    18          73          73     -9.68361     -4.03725     -0.24096     10.49428      0.00000
   55  (g)           I   12       21    15          73          73     -0.02970     -0.14691      0.54657      0.56675      0.00000
   56  (g)           I   12       21    25          73          73     -0.12274      1.39016      3.91159      4.15309      0.00000
   57  (g)           I   12       21    24          73          73     -1.25460      0.15914      3.44828      3.67287      0.00000
   58  (g)           I   12       21    23          73          73     -0.12619      0.81281      2.38321      2.52117      0.00000
   59  (g)           I   12       21    27          73          73      1.39939      5.08227      7.51962      9.18327      0.00000
   60  (u~)          V   11       -2    26          73          73      2.89930      7.50857      8.72410     11.86990      0.00560
   61  (u)           A   12        2    31          91          91      2.06978     -3.05418     -5.61577      6.71930      0.00560
   62  (g)           I   12       21    44          91          91     -0.23835      0.13849      1.08896      1.12330      0.00000
   63  (g)           I   12       21    43          91          91     -2.15456     -0.80571      0.13785      2.30441      0.00000
   64  (g)           I   12       21    50          91          91      2.92351      0.47380      2.41516      3.82157      0.00000
   65  (g)           I   12       21    49          91          91     14.75801      5.72986     14.60889     21.54181      0.00000
   66  (g)           I   12       21    41          91          91      6.51083      2.61590      3.06918      7.65857      0.00000
   67  (g)           I   12       21    48          91          91     11.81774     -6.56210    -18.85316     23.19831      0.00000
   68  (g)           I   12       21    47          91          91      1.41318     -1.20728     -3.26543      3.75734      0.00000
   69  (g)           I   12       21    46          91          91      1.58193     -0.22443     -2.48869      2.95744      0.00000
   70  (g)           I   12       21    45          91          91      9.54865     -1.43393     -9.90045     13.82938      0.00000
   71  (g)           I   12       21    38          91          91      2.25644      0.88705     -1.73723      2.98268      0.00000
   72  (d~)          V   11       -1    37          91          91      7.66346      2.82714    -12.47235     14.90909      0.00990
   73  (string)          11       92    51          74          90    -58.15062      0.61539     33.01305     95.19680     67.75468
   74  (eta)             11      221    73         123         125    -10.24036     -1.30218      1.37509     10.42839      0.54750
   75  n0                 1     2112    73           0           0    -18.93694     -4.18797      2.49748     19.57721      0.93960
   76  p~-                1    -2212    73           0           0    -14.20916     -3.08438      1.25868     14.62458      0.93830
   77  (rho0)            11      113    73         126         127     -6.22881     -2.31614      1.02168      6.76712      0.76646
   78  K+                 1      321    73           0           0     -3.92215     -0.97100      0.06270      4.07107      0.49360
   79  (Lambda0)         11     3122    73         128         129     -3.01880     -0.44703      0.30808      3.26381      1.11560
   80  n~0                1    -2112    73           0           0     -3.75048     -1.33794      0.39685      4.11053      0.93960
   81  (eta')            11      331    73         130         131     -0.12766      0.40474      0.85328      1.35133      0.95807
   82  (pi0)             11      111    73         132         133     -0.50601     -0.14221      0.41852      0.68531      0.13500
   83  (omega)           11      223    73         134         136     -0.63008      0.00890      3.96938      4.09435      0.78145
   84  (pi0)             11      111    73         137         138      0.12911      0.68122      0.54176      0.89020      0.13500
   85  (rho0)            11      113    73         139         140      0.01639      1.31382      3.53153      3.84545      0.76771
   86  pi-                1     -211    73           0           0     -0.25161     -0.09021      0.25972      0.39798      0.13960
   87  (rho+)            11      213    73         141         142      0.28192      1.27522      2.20383      2.67551      0.77191
   88  pi-                1     -211    73           0           0     -0.03182      1.50185      1.71206      2.28193      0.13960
   89  K+                 1      321    73           0           0      1.10097      1.59131      2.64052      3.31065      0.49360
   90  (K*-)             11     -323    73         143         144      2.17486      7.71740      9.96190     12.82138      0.92739
   91  (string)          11       92    61          92         122     58.15061     -0.61540    -33.01305    104.80321     80.69683
   92  (rho+)            11      213    91         145         146      1.55024     -2.13893     -3.14091      4.17312      0.75586
   93  pi-                1     -211    91           0           0     -0.17159      0.04685      0.08614      0.24196      0.13960
   94  pi+                1      211    91           0           0      0.24570     -0.03059     -0.39585      0.48733      0.13960
   95  (rho0)            11      113    91         147         148     -0.05235      0.18826     -0.45908      0.89626      0.74455
   96  (rho0)            11      113    91         149         150     -0.32546     -0.90485     -0.78443      1.47577      0.79869
   97  (rho-)            11     -213    91         151         152     -0.25670     -0.63219     -0.01561      1.05392      0.80308
   98  (Delta++)         11     2224    91         153         154      0.88891      0.12972      0.83826      1.74275      1.23593
   99  K-                 1     -321    91           0           0     -0.03388      0.33010      0.05701      0.59750      0.49360
  100  (Sigma~-)         11    -3222    91         155         156      0.07562     -0.53240      0.50627      1.40005      1.18940
  101  p+                 1     2212    91           0           0      3.33352      2.05592      4.01892      5.68958      0.93830
  102  (rho+)            11      213    91         157         158      1.90521      0.28713      1.53127      2.54161      0.63461
  103  p~-                1    -2212    91           0           0      1.36015      0.95944      1.70996      2.56416      0.93830
  104  (eta)             11      221    91         159         161      5.08637      0.98859      3.62885      6.34955      0.54750
  105  (rho-)            11     -213    91         162         163      4.25581      2.04714      3.53767      5.93431      0.63108
  106  (rho+)            11      213    91         164         165      2.21954      0.96066      3.02820      3.94127      0.71724
  107  (eta)             11      221    91         166         168      1.47892      0.23889      0.82782      1.79702      0.54750
  108  (pi0)             11      111    91         169         170      1.98223      0.39680      0.15178      2.03174      0.13500
  109  (rho-)            11     -213    91         171         172      0.67407      0.00155     -1.27522      1.61024      0.71575
  110  (rho+)            11      213    91         173         174      3.49583     -1.53409     -4.96288      6.33350      0.95327
  111  (Delta0)          11     2114    91         175         176      1.66564     -1.34819     -2.37829      3.39981      1.14476
  112  (eta)             11      221    91         177         179      0.86265     -0.26884     -1.72434      2.02226      0.54750
  113  (Delta~-)         11    -2214    91         180         181      3.12720     -1.49499     -4.64393      5.91198      1.17091
  114  (eta)             11      221    91         182         183      4.40704     -1.97240     -6.96861      8.49551      0.54750
  115  pi+                1      211    91           0           0      0.73868     -0.26451     -0.84399      1.16078      0.13960
  116  (pi0)             11      111    91         184         185      0.19280     -0.10577     -0.69190      0.73845      0.13500
  117  (rho-)            11     -213    91         186         187      3.50644     -0.39157     -3.81033      5.24824      0.75957
  118  (Sigma*+)         11     3224    91         188         189      3.62885     -0.37994     -4.27561      5.78618      1.37332
  119  (Xi*~0)           11    -3324    91         190         191      4.58324      0.53697     -5.60459      7.42030      1.53458
  120  K-                 1     -321    91           0           0      4.83667      1.54290     -6.58041      8.32582      0.49360
  121  (rho0)            11      113    91         192         193      0.81491      0.29522     -1.28553      1.68575      0.66175
  122  pi+                1      211    91           0           0      2.07437      0.37776     -3.09369      3.74648      0.13960
  123  (pi0)             11      111    74         194         195     -2.10779     -0.28820      0.21802      2.14280      0.13500
  124  (pi0)             11      111    74         196         197     -2.68848     -0.44180      0.33111      2.74790      0.13500
  125  (pi0)             11      111    74         198         199     -5.44409     -0.57218      0.82596      5.53769      0.13500
  126  pi+                1      211    77           0           0     -4.66494     -1.98637      0.59457      5.10689      0.13960
  127  pi-                1     -211    77           0           0     -1.56387     -0.32977      0.42711      1.66022      0.13960
  128  n0                 1     2112    79           0           0     -2.67232     -0.46645      0.20868      2.87841      0.93960
  129  (pi0)             11      111    79         200         201     -0.34648      0.01942      0.09940      0.38540      0.13500
  130  gamma              1       22    81           0           0      0.15036      0.00694      0.20410      0.25359      0.00000
  131  (rho0)            11      113    81         202         203     -0.27802      0.39780      0.64919      1.09774      0.74030
  132  gamma              1       22    82           0           0     -0.37766     -0.11643      0.38142      0.54924      0.00000
  133  gamma              1       22    82           0           0     -0.12835     -0.02578      0.03709      0.13607      0.00000
  134  pi-                1     -211    83           0           0     -0.55143     -0.02633      2.19196      2.26471      0.13960
  135  pi+                1      211    83           0           0      0.03016     -0.02559      0.23412      0.27544      0.13960
  136  (pi0)             11      111    83         204         205     -0.10881      0.06083      1.54330      1.55420      0.13500
  137  gamma              1       22    84           0           0      0.10429      0.28346      0.17781      0.35049      0.00000
  138  gamma              1       22    84           0           0      0.02482      0.39776      0.36395      0.53971      0.00000
  139  pi-                1     -211    85           0           0     -0.02925      0.81104      1.19718      1.45305      0.13960
  140  pi+                1      211    85           0           0      0.04564      0.50278      2.33435      2.39240      0.13960
  141  pi+                1      211    87           0           0     -0.10049      0.97231      1.66839      1.93869      0.13960
  142  (pi0)             11      111    87         206         207      0.38241      0.30290      0.53544      0.73682      0.13500
  143  (K~0)             11     -311    90         208         208      1.63227      5.30466      6.44776      8.52203      0.49770
  144  pi-                1     -211    90           0           0      0.54259      2.41274      3.51414      4.29935      0.13960
  145  pi+                1      211    92           0           0      0.37519     -0.42730     -1.16477      1.30366      0.13960
  146  (pi0)             11      111    92         209         210      1.17505     -1.71164     -1.97614      2.86946      0.13500
  147  pi-                1     -211    95           0           0     -0.02511     -0.25130     -0.27599      0.39931      0.13960
  148  pi+                1      211    95           0           0     -0.02724      0.43956     -0.18308      0.49696      0.13960
  149  pi+                1      211    96           0           0      0.01872     -0.34856     -0.70919      0.80267      0.13960
  150  pi-                1     -211    96           0           0     -0.34419     -0.55629     -0.07524      0.67311      0.13960
  151  pi-                1     -211    97           0           0     -0.31262     -0.28175     -0.33663      0.55671      0.13960
  152  (pi0)             11      111    97         211         212      0.05592     -0.35044      0.32102      0.49721      0.13500
  153  p+                 1     2212    98           0           0      0.71834     -0.11831      0.72858      1.39328      0.93830
  154  pi+                1      211    98           0           0      0.17057      0.24802      0.10968      0.34947      0.13960
  155  n~0                1    -2112   100           0           0      0.16668     -0.28801      0.46995      1.10201      0.93960
  156  pi-                1     -211   100           0           0     -0.09107     -0.24439      0.03632      0.29804      0.13960
  157  pi+                1      211   102           0           0      1.74927      0.13762      1.23164      2.14833      0.13960
  158  (pi0)             11      111   102         213         214      0.15594      0.14950      0.29963      0.39328      0.13500
  159  pi+                1      211   104           0           0      1.81616      0.41035      1.28781      2.26821      0.13960
  160  pi-                1     -211   104           0           0      2.55951      0.39048      1.74498      3.12538      0.13960
  161  (pi0)             11      111   104         215         216      0.71070      0.18776      0.59605      0.95596      0.13500
  162  pi-                1     -211   105           0           0      3.62668      1.80466      3.22112      5.17733      0.13960
  163  (pi0)             11      111   105         217         218      0.62913      0.24248      0.31655      0.75699      0.13500
  164  pi+                1      211   106           0           0      2.01360      0.65080      2.44717      3.23825      0.13960
  165  (pi0)             11      111   106         219         220      0.20593      0.30985      0.58104      0.70303      0.13500
  166  (pi0)             11      111   107         221         222      0.68124      0.15877      0.32632      0.78359      0.13500
  167  (pi0)             11      111   107         223         224      0.15443      0.08805      0.11955      0.25322      0.13500
  168  (pi0)             11      111   107         225         226      0.64325     -0.00793      0.38194      0.76022      0.13500
  169  gamma              1       22   108           0           0      0.52260      0.07609      0.09222      0.53610      0.00000
  170  gamma              1       22   108           0           0      1.45963      0.32071      0.05956      1.49564      0.00000
  171  pi-                1     -211   109           0           0      0.17198     -0.28518     -0.30458      0.47240      0.13960
  172  (pi0)             11      111   109         227         228      0.50209      0.28673     -0.97064      1.13784      0.13500
  173  pi+                1      211   110           0           0      2.93956     -1.28882     -3.54111      4.78132      0.13960
  174  (pi0)             11      111   110         229         230      0.55627     -0.24528     -1.42177      1.55218      0.13500
  175  n0                 1     2112   111           0           0      1.56887     -1.33211     -2.20490      3.15916      0.93960
  176  (pi0)             11      111   111         231         232      0.09677     -0.01608     -0.17339      0.24065      0.13500
  177  gamma              1       22   112           0           0      0.01295      0.04499     -0.05018      0.06863      0.00000
  178  pi+                1      211   112           0           0      0.55623     -0.35914     -1.18580      1.36528      0.13960
  179  pi-                1     -211   112           0           0      0.29347      0.04532     -0.48835      0.58835      0.13960
  180  p~-                1    -2212   113           0           0      2.40690     -1.27183     -3.45434      4.49706      0.93830
  181  (pi0)             11      111   113         233         234      0.72030     -0.22316     -1.18960      1.41492      0.13500
  182  gamma              1       22   114           0           0      3.72994     -1.81174     -6.19765      7.45692      0.00000
  183  gamma              1       22   114           0           0      0.67710     -0.16066     -0.77096      1.03858      0.00000
  184  gamma              1       22   116           0           0      0.09058     -0.06419     -0.15700      0.19229      0.00000
  185  gamma              1       22   116           0           0      0.10222     -0.04158     -0.53489      0.54616      0.00000
  186  pi-                1     -211   117           0           0      0.43395     -0.27180     -0.52830      0.74885      0.13960
  187  (pi0)             11      111   117         235         236      3.07249     -0.11977     -3.28203      4.49939      0.13500
  188  (Lambda0)         11     3122   118         237         238      3.42267     -0.30045     -3.79719      5.24101      1.11560
  189  pi+                1      211   118           0           0      0.20618     -0.07949     -0.47842      0.54517      0.13960
  190  (Xi~+)            11    -3312   119         239         240      4.23992      0.59179     -5.04930      6.75043      1.32130
  191  pi-                1     -211   119           0           0      0.34332     -0.05482     -0.55529      0.66986      0.13960
  192  pi+                1      211   121           0           0      0.34304      0.26564     -1.03898      1.13455      0.13960
  193  pi-                1     -211   121           0           0      0.47188      0.02957     -0.24655      0.55120      0.13960
  194  gamma              1       22   123           0           0     -0.59562     -0.14153      0.07625      0.61694      0.00000
  195  gamma              1       22   123           0           0     -1.51217     -0.14667      0.14177      1.52587      0.00000
  196  gamma              1       22   124           0           0     -1.59588     -0.32397      0.22387      1.64375      0.00000
  197  gamma              1       22   124           0           0     -1.09260     -0.11784      0.10724      1.10415      0.00000
  198  gamma              1       22   125           0           0     -1.44943     -0.10471      0.25587      1.47557      0.00000
  199  gamma              1       22   125           0           0     -3.99466     -0.46746      0.57009      4.06212      0.00000
  200  gamma              1       22   129           0           0     -0.26109      0.07007      0.05160      0.27521      0.00000
  201  gamma              1       22   129           0           0     -0.08540     -0.05065      0.04780      0.11019      0.00000
  202  pi+                1      211   131           0           0     -0.24040      0.45140      0.11596      0.54266      0.13960
  203  pi-                1     -211   131           0           0     -0.03762     -0.05360      0.53323      0.55507      0.13960
  204  gamma              1       22   136           0           0     -0.04084      0.06750      1.27167      1.27412      0.00000
  205  gamma              1       22   136           0           0     -0.06797     -0.00668      0.27163      0.28008      0.00000
  206  gamma              1       22   142           0           0      0.12366      0.02444      0.13402      0.18398      0.00000
  207  gamma              1       22   142           0           0      0.25876      0.27847      0.40142      0.55284      0.00000
  208  K_L0               1      130   143           0           0      1.63227      5.30466      6.44776      8.52203      0.49770
  209  gamma              1       22   146           0           0      0.72737     -1.07804     -1.14781      1.73456      0.00000
  210  gamma              1       22   146           0           0      0.44768     -0.63360     -0.82833      1.13490      0.00000
  211  gamma              1       22   152           0           0      0.05817     -0.13211      0.05850      0.15575      0.00000
  212  gamma              1       22   152           0           0     -0.00225     -0.21834      0.26252      0.34146      0.00000
  213  gamma              1       22   158           0           0      0.00265     -0.02338      0.03177      0.03954      0.00000
  214  gamma              1       22   158           0           0      0.15328      0.17289      0.26786      0.35374      0.00000
  215  gamma              1       22   161           0           0      0.13516      0.03401      0.18738      0.23353      0.00000
  216  gamma              1       22   161           0           0      0.57554      0.15374      0.40867      0.72242      0.00000
  217  gamma              1       22   163           0           0      0.30126      0.04572      0.12559      0.32958      0.00000
  218  gamma              1       22   163           0           0      0.32786      0.19676      0.19096      0.42741      0.00000
  219  gamma              1       22   165           0           0      0.09395      0.05367      0.23060      0.25473      0.00000
  220  gamma              1       22   165           0           0      0.11198      0.25618      0.35043      0.44830      0.00000
  221  gamma              1       22   166           0           0      0.49958      0.11652      0.30257      0.59557      0.00000
  222  gamma              1       22   166           0           0      0.18167      0.04226      0.02376      0.18802      0.00000
  223  gamma              1       22   167           0           0     -0.02210     -0.00142      0.00028      0.02215      0.00000
  224  gamma              1       22   167           0           0      0.17653      0.08947      0.11927      0.23107      0.00000
  225  gamma              1       22   168           0           0      0.08815      0.01534      0.10957      0.14146      0.00000
  226  gamma              1       22   168           0           0      0.55510     -0.02327      0.27237      0.61876      0.00000
  227  gamma              1       22   172           0           0      0.12712      0.11287     -0.19308      0.25725      0.00000
  228  gamma              1       22   172           0           0      0.37498      0.17386     -0.77756      0.88059      0.00000
  229  gamma              1       22   174           0           0      0.28732     -0.10621     -0.87647      0.92846      0.00000
  230  gamma              1       22   174           0           0      0.26895     -0.13907     -0.54530      0.62372      0.00000
  231  gamma              1       22   176           0           0      0.12088     -0.04281     -0.11611      0.17299      0.00000
  232  gamma              1       22   176           0           0     -0.02411      0.02673     -0.05729      0.06766      0.00000
  233  gamma              1       22   181           0           0      0.37690     -0.05568     -0.65663      0.75915      0.00000
  234  gamma              1       22   181           0           0      0.34340     -0.16748     -0.53297      0.65577      0.00000
  235  gamma              1       22   187           0           0      0.08510     -0.01401     -0.06558      0.10835      0.00000
  236  gamma              1       22   187           0           0      2.98739     -0.10576     -3.21645      4.39104      0.00000
  237  p+                 1     2212   188           0           0      2.76880     -0.33055     -3.14146      4.30403      0.93830
  238  pi-                1     -211   188           0           0      0.65387      0.03010     -0.65573      0.93698      0.13960
  239  (Lambda~0)        11    -3122   190         241         242      3.97897      0.50910     -4.80418      6.35736      1.11560
  240  pi+                1      211   190           0           0      0.26095      0.08270     -0.24512      0.39307      0.13960
  241  p~-                1    -2212   239           0           0      3.53794      0.36931     -4.30069      5.65949      0.93830
  242  pi+                1      211   239           0           0      0.44102      0.13979     -0.50348      0.69787      0.13960
                   sum charge:  0.00   sum momentum and inv. mass:     -0.00001

   ..... skipped ....
\end{verbatim}
}
{
\small
\renewcommand{\baselinestretch}{0.4}
\begin{verbatim}
 ***************************************************************************
 *                      KarLud  FINAL  REPORT                              *
 *                          window A                                       *
 *             4537                 total no of events       NEVTOT     A0 *
 *                0                 WT<0        events       NEVNEG     A1 *
 *      19.33542764                 xs_cru VESKO [pb]        XCVESK     A2 *
 *      19.17361574  +- 0.01754039  xs_est VESKO [pb]        XSVE       A3 *
 *      19.19161478  +- 0.00001919  xs_est Gauss [pb]        XSGS       A4 *
 *      -0.00093786  +- 0.00091582  XCVE/XCGS-1                         A5 *
 *     251.14598768  +- 0.24820021      <WTKARL>             WTKARL     A6 *
 ***************************************************************************


 ***************************************************************************
 *                        KarFin output - window A                         *
 *                            Weight Statistics                            *
 *       0.99643450  +- 0.00368336   general weight          WT         A1 *
 *             4263                  generated events        NEVGEN     A2 *
 *       0.08139808                  aver. ph. multi.        AVMULT     A3 *
 *                0                  Marked photons          MARTOT     A4 *
 *       0.99108609  +- 0.00145252   Kinematics, smin        WT1        A5 *
 *       0.99999743  +- 0.00000168   Jacobian                WT2        A6 *
 *       1.00534843  +- 0.00335486   Photon ang. dist.       WT3        A7 *
 ***************************************************************************

 
 ***************************************************************************
 *                        KarFin output - window B                         *
 *                            on mass weights                              *
 *       0.96568167  +- 0.00245351   removal wgt WTREM         WT6      B1 *
 *             4263                  no. of raw events                  B2 *
 *                0                  WT6=0      events                  B3 *
 *       0.08139808                  raw ph. multipl.                   B4 *
 *       0.99951442  +- 0.00245437   control wgt WCTRL         WT5      B5 *
 *   0.10000000E-05                  epsilon                            B6 *
 *   0.10000000E-08                  delta                              B7 *
 ***************************************************************************



 ***************************************************************************
 *                         ******** DADMEL FINAL REPORT  ********          *
 *                  48     NEVRAW = NO. OF EL  DECAYS TOTAL                *
 *                  18     NEVACC = NO. OF EL   DECS. ACCEPTED             *
 *                   0     NEVOVR = NO. OF OVERWEIGHTED EVENTS             *
 *         0.48927E-12     PARTIAL WTDTH ( ELECTRON) IN GEV UNITS          *
 *         1.207993507     IN UNITS GFERMI**2*MASS**5/192/PI**3            *
 *         0.128391966     RELATIVE ERROR OF PARTIAL WIDTH                 *
 *                         COMPLETE QED CORRECTIONS INCLUDED               *
 *                         BUT ONLY V-A CUPLINGS                           *
 ***************************************************************************

   ..... skipped ....

 ***************************************************************************
 *                         *****TAUOLA LIBRARY: VERSION 2.6 ******         *
 *                         ***********August   1995***************         *
 *                         **AUTHORS: S.JADACH, Z.WAS*************         *
 *                         **R. DECKER, M. JEZABEK, J.H.KUEHN*****         *
 *                         **AVAILABLE FROM: WASM AT CERNVM ******         *
 *                         ***** PUBLISHED IN COMP. PHYS. COMM.***         *
 *                         *******CERN-TH-5856 SEPTEMBER 1990*****         *
 *                         *******CERN-TH-6195 SEPTEMBER 1991*****         *
 *                         *******CERN-TH-6793 NOVEMBER  1992*****         *
 *                         ******DEXAY ROUTINE: FINAL REPORT******         *
 *                 105     NEV1   = NO. OF TAU+ DECS. ACCEPTED             *
 *                  91     NEV2   = NO. OF TAU- DECS. ACCEPTED             *
 *                 196     NEVTOT = SUM                                    *
 *    NOEVTS  PART.WIDTH     ERROR       ROUTINE    DECAY MODE             *
 *         9   1.2079935   0.1283920     DADMEL     ELECTRON               *
 *         4   0.8730308   0.2060235     DADMMU     MUON                   *
 *         6   0.6106793   0.0000000     DADMPI     PION                   *
 *         9   1.2446948   0.0659817     DADMRO     RHO (->2PI)            *
 *         9   0.7094790   0.1167541     DADMAA     A1  (->3PI)            *
 *        10   0.0400228   0.0000000     DADMKK     KAON                   *
 *        10   0.0698563   0.0652813     DADMKS     K*                     *
 *         7   0.0798995   0.1472948  TAU-  --> 2PI-,  PI0,  PI+           *
 *        12   0.0178338   0.0767882  TAU-  --> 3PI0,        PI-           *
 *        13   0.0580654   0.1190726  TAU-  --> 2PI-,  PI+, 2PI0           *
 *        11   0.0267028   0.1478714  TAU-  --> 3PI-, 2PI+,                *
 *         8   0.0042604   0.1847376  TAU-  --> 3PI-, 2PI+,  PI0           *
 *         6   0.0029619   0.2054046  TAU-  --> 2PI-,  PI+, 3PI0           *
 *         9   0.0063946   0.1197646  TAU-  -->  K-, PI-,  K+              *
 *         7   0.0051756   0.2002329  TAU-  -->  K0, PI-, K0B              *
 *        11   0.0005735   0.1100686  TAU-  -->  K-,  K0, PI0              *
 *         8   0.0059857   0.1026906  TAU-  --> PI0, PI0,  K-              *
 *        11   0.0325313   0.0894969  TAU-  -->  K-, PI-, PI+              *
 *        10   0.0291474   0.1199303  TAU-  --> PI-, K0B, PI0              *
 *         9   0.0106989   0.1192819  TAU-  --> ETA, PI-, PI0              *
 *         7   0.0028271   0.0911550  TAU-  --> PI-, PI0, GAM              *
 *        10   0.0172013   0.1098466  TAU-  -->  K-,  K0                   *
 *                    THE ERROR IS RELATIVE AND  PART.WIDTH                *
 *                    IN UNITS GFERMI**2*MASS**5/192/PI**3                 *
 ***************************************************************************





 ***************************************************************************
 *                          YFSWW3  Final  Report                          *
 *                                Window A                                 *
 *                                                                         *
 *     200.00000000                  CMS energy total        CMSENE     A0 *
 *      16.96445667  +- 0.24052939   xs_tot [pb]             XSMCPB     A1 *
 *       0.01417843                  relative error          EREL       A2 *
 *             4537                  total no of events      NEVTOT     A3 *
 *             1000                  accepted    events      NEVACC     A4 *
 *                0                  WT<0        events      NEVNEG     A5 *
 *       0.00000000  +- 0.00000000   xsec/xtot: WT<0         XSNEG      A6 *
 *               17                  WT>WTMAX    events      NEVOVE     A7 *
 *       0.00096638  +- 0.00028479   xsec/xtot:WT>WTMAX      XSOVE      A8 *
 *       4.00000000                  WTMAX                   WTMAX      A9 *
 *       0.87737685  +- 0.01243983       <WT>                AWTOT     A10 *
 ***************************************************************************


 ***************************************************************************
 *                          YFSWW3  final  report                          *
 *                                Window B:                                *
 *                                   ISR                                   *
 *                                                                         *
 *      16.71805108  +- 0.20216719       xsec total         O(alf0)     B1 *
 *      17.20746824  +- 0.21080089       xsec total         O(alf1)     B2 *
 *      17.22391080  +- 0.21107044       xsec total         O(alf2)     B3 *
 *      17.22440793  +- 0.21107682       xsec total         O(alf3)     B4 *
 *      16.71805108  +- 0.20216719       xsec(beta00)       O(alf0)     B5 *
 *      17.67935928  +- 0.21379205       xsec(beta01)       O(alf1)     B6 *
 *      -0.47189104  +--0.01736687       xsec(beta10)       O(alf1)     B7 *
 *      17.70699748  +- 0.21412627       xsec(beta02)       O(alf2)     B8 *
 *      -0.48533564  +--0.01785827       xsec(beta11)       O(alf2)     B9 *
 *       0.00224896  +- 0.00031913       xsec(beta20)       O(alf2)    B10 *
 *      17.70752722  +- 0.00002720       xsec(beta03)       O(alf3)    B11 *
 *      -0.48528352  +- 0.00008275       xsec(beta12)       O(alf3)    B12 *
 *       0.00216423  +- 0.00033020       xsec(beta21)       O(alf3)    B13 *
 *       0.00000000  +- 0.01265929       xsec(beta30)       O(alf3)    B14 *
 *                           xsec_tot differences                          *
 *       0.48941716  +- 0.01874270        xstot(alf1-0)     O(alf1)    B15 *
 *       0.01644256  +- 0.00052593        xstot(alf2-1)     O(alf2)    B16 *
 *       0.00049713  +- 0.00001292        xstot(alf3-2)     O(alf3)    B17 *
 *                              betas differences                          *
 *       0.96130820  +- 0.01162486        xs(beta01-00)     O(alf1)    B18 *
 *      -0.47189104  +--0.01736687        xs(beta10)        O(alf1)    B19 *
 *       0.02763819  +- 0.00033422        xs(beta02-01)     O(alf2)    B20 *
 *      -0.01344459  +--0.00049160        xs(beta11-10)     O(alf2)    B21 *
 *       0.00224896  +- 0.00031913        xs(beta20)        O(alf2)    B22 *
 *       0.00052974  +- 0.00000641        xs(beta03-02)     O(alf3)    B23 *
 *       0.00005212  +- 0.00000324        xs(beta12-11)     O(alf3)    B24 *
 *      -0.00008473  +--0.00001141        xs(beta21-20)     O(alf3)    B25 *
 *       0.00000000  +- 0.01265929        xs(beta30)        O(alf3)    B26 *
 ***************************************************************************


 ***************************************************************************
 *                          YFSWW3  final  report                          *
 *                                Window C:                                *
 *                        O(alf1) EWRC in WW-prod.                         *
 *                                                                         *
 *      16.94721804  +- 0.24020843       xs_tot: EW-ex      O(alf1)     C1 *
 *      16.92711416  +- 0.24008944       xs_tot: EW-ap      O(alf1)     C2 *
 *      17.20746824  +- 0.21080089       xs_tot: ISR-LL     O(alf1)     C3 *
 *                           xsec_tot differences                          *
 *       0.02010388  +- 0.00091149       EW-ex - EW-ap      O(alf1)     C4 *
 *      -0.26025020  +--0.11286284       EW-ex - ISR-LL     O(alf1)     C5 *
 *                              betas differences                          *
 *       0.02010388  +- 0.00091149       bet01(ex - ap)     O(alf1)     C6 *
 *      -0.40008677  +--0.00639368       bet01(ex - LL)     O(alf1)     C7 *
 *      -0.02807084  +--0.00454222       bet10(ex - LL)     O(alf1)     C8 *
 ***************************************************************************



   ..... skipped ....

                                                            
                              ============ demo ============
       16.96445667  +- 0.24052939          MC Best, XPAR, YFSWW3
       18.96537427  +- 0.00085125          SemiAnal Born, KorWan
       17.42293229  +- 0.00255375 SemiAnal O(alf3)exp.LL, KorWan
                              ========== End demo ==========
\end{verbatim}
}

\newpage


\end{document}